\documentclass[preprint,10pt]{elsarticle}

\usepackage{graphics}
\usepackage{amssymb}
\usepackage{amsmath, amsthm}
\usepackage{mathrsfs}

\journal{Journal of Computational Physics}

\begin{document}

\begin{frontmatter}

%% Title, authors and addresses

%% use the tnoteref command within \title for footnotes;
%% use the tnotetext command for theassociated footnote;
%% use the fnref command within \author or \address for footnotes;
%% use the fntext command for theassociated footnote;
%% use the corref command within \author for corresponding author footnotes;
%% use the cortext command for theassociated footnote;
%% use the ead command for the email address,
%% and the form \ead[url] for the home page:
%% \title{Title\tnoteref{label1}}
%% \tnotetext[label1]{}
%% \author{Name\corref{cor1}\fnref{label2}}
%% \ead{email address}
%% \ead[url]{home page}
%% \fntext[label2]{}
%% \cortext[cor1]{}
%% \address{Address\fnref{label3}}
%% \fntext[label3]{}

\title{ An iterative, dynamically stabilized method of data unfolding }

%% use optional labels to link authors explicitly to addresses:
%% \author[label1,label2]{}
%% \address[label1]{}
%% \address[label2]{}

\author{ Bogdan MALAESCU }

\address{Laboratoire de l'Acc{\'e}l{\'e}rateur Lin{\'e}aire,
        IN2P3/CNRS et Universit\'e Paris-Sud 11 (UMR 8607), F--91405, Orsay Cedex, France}

\begin{abstract}
%%Text of abstract
We propose a new iterative unfolding method for experimental data, making use of a regularization function.
The use of this function allows one to build an improved normalization procedure for Monte Carlo spectra, unbiased by 
the presence of possible new structures in data.
We are able to unfold, in a dynamically stable way, data spectra which can be strongly affected by fluctuations in the 
background subtraction and simultaneously reconstruct structures which were not initially simulated.
This method also allows one to control the amount of correlations introduced between the bins of the unfolded spectrum, when
the transfers of events correcting the systematic detector effects are performed.
\end{abstract}

\begin{keyword}
%% keywords here, in the form: keyword \sep keyword
unfolding \sep iterations
%% PACS codes here, in the form: \PACS code \sep code

%% MSC codes here, in the form: \MSC code \sep code
%% or \MSC[2008] code \sep code (2000 is the default)

\begin{flushright}
\normalsize
LAL 09-107 \\
\end{flushright}

\end{keyword}

\end{frontmatter}

%% \linenumbers

%% main text
\section{ Introduction }
\label{SecIntroduction}
Experimental distributions in high-energy physics are altered by detector effects.
This can be due to limited acceptance, finite resolution, or other systematic effects
producing a transfer of events between different regions of the spectra.
Provided that they are well controlled experimentally, all these effects can be included 
in the Monte Carlo simulations (MC) of the detector response.
One can then use these simulations to correct the data for the same effects.
Several deconvolution methods for data affected by detector effects were described in the
past (see for example \cite{Hocker:1995kb,Blobel:2002pu,Kondor:1982ah, 
Lindemann:1995ut,D'Agostini:1994zf,Acton:1993zh}).

The aim of this paper is to illustrate a new unfolding method allowing one to obtain a data
distribution as close as possible to the ``real'' one for rather difficult, yet realistic, 
examples.
This method is based on the idea that if the MC simulation provides a relatively good 
description of the data and of the detector response, one can use the transfer matrix to
compute a matrix of unfolding probabilities.
If the first condition is not fulfilled one can iteratively improve the transfer matrix.
Our method is using a first step, providing a good result if the difference 
between data and normalized reconstructed MC is relatively small, on the entire spectrum.
If this is not the case, one should proceed with a series of iterations.

The author learned the existence of the iterative methods present in the above list, 
when this study was close to completion.
Two of them (\cite{Kondor:1982ah,Lindemann:1995ut}) use a direct comparison 
of data and reconstructed MC events to compute normalization coefficients for the true MC 
events.
However, it has been shown in \cite{Multhei:1985qs,Multhei:1986ps} that this approach 
has a rather slow convergence.
Further on, in \cite{D'Agostini:1994zf,Acton:1993zh} one can find two methods 
based on the comparison of the true MC with the unfolding result at a given iteration.
This comparison allows the computation of the probability matrix to be used at the next step.
For the parts of our procedure which are similar to the ones in these last two papers,
we will emphasize which are the differences that make our method more general and better behaved.

In Section \ref{sec:ImportantConsiderations} we present a series of considerations taken into account
in the unfolding method developped in the present paper.
We describe the ingredients of the method in Sections \ref{Sec:RegF} - \ref{Sec:FoldAndUnf}.
In Section \ref{Sec:IterUnfStrategy}, we assemble all these elements into an unfolding strategy.
We provide two examples of applications of the method, with some possible tests, in 
Sections \ref{Sec:ComplexEx} and \ref{Sec:SimpleEx}.
We introduce the notations systematically and we also recall them in Appendix \ref{App:Notation}.

\section{Important considerations for the design of the procedure}
\label{sec:ImportantConsiderations}
In the unfolding procedure we will not concentrate on the correction of acceptance effects.
It is straightforward to perform it on the distribution corrected for the effects resulting 
in a transfer of events between different bins of the spectrum.
Actually, these transfers have a physical meaning only for events which are in the acceptance 
of the detector and which passed all the cuts.
The method described in \cite{D'Agostini:1994zf} computes at each iteration the spectrum 
corrected for the acceptance.
This result is then used to improve the folding matrix (i.e. the probabilities of the causes,
as they are called in this reference).
Furthermore, the folding probabilities for each bin are normalized to the corresponding 
acceptance.
It is straightforward to prove analytically that this method is completely equivalent 
to the one used in \cite{Acton:1993zh}, where the acceptance is ignored in the unfolding and 
the corresponding correction is done at the end.

A first, rather tricky point is the way how the unfolding deals with new structures, absent in the 
MC, but which can be present in the data.
These structures are affected by the detector effects, and hence they need to be corrected.
It seems that the Singular Values Decomposition (SVD) \cite{Hocker:1995kb} and the iterative \cite{Blobel:2002pu,
Kondor:1982ah,D'Agostini:1994zf,Acton:1993zh} methods provide a natural way of performing this correction.
If the new structures in the data contain a relatively important number of events, they could also
affect the normalization of MC spectra with respect to the data.
The unfolding procedure described here makes use of a comparison method between data and MC spectra, 
which ignores significant shape differences when computing the relative normalization factor.
This allows a coherent comparison of data and MC spectra and improves the convergence of the algorithm 
when the data contain structures which are not present in the MC simulation.

It is also important to keep in mind that experimental spectra are generally obtained after 
background subtraction.
This operation results in an increase of errors for the corresponding data points.
Due to bin-to-bin or correlated fluctuations of the subtracted background, these points can 
fluctuate within their errors, or they can hide systematic effects as large as the error 
components due to background subtraction.
When computing the central values of the final distribution, the unfolding procedure 
has to take into account the size of the experimental errors, including the ones from background subtraction.
Not doing so could result in too large a transfer of events from/to different regions of the 
spectrum.
This would affect the regions of important background subtraction, but could also bias 
the central values of more precise points in another region.
Such a systematic effect of the procedure is to be avoided. 
To our knowledge, none of the previous methods aim at dealing with this second type of problem,
and at distinguishing it from the first one
\footnote{In \cite{D'Agostini:1994zf} a proposal was made to deal with the background 
subtraction in the unfolding procedure itself, by introducing an additional ``cause
responsible for the observables''. 
This would indeed perform the subtraction, but would not solve the problem
of the systematic effects and/or fluctuations by itself.}.

We are going to describe a method which is be able to deal with the potential problems 
described above.
The fluctuations which can be generated when using this type of method are dynamically 
reduced, in the sense that the way in which bin-by-bin corrections are performed avoids 
this problem.
Therefore, no additional smoothing by a model-dependent fitting of the unfolding result 
at each iteration step (as proposed in \cite{D'Agostini:1994zf}), is required.

The transfer of events performed by the unfolding introduces bin-to-bin correlations for the final
spectrum.
For further use of the data, these correlations are to be reduced as much as possible, 
if this can be done without introducing significant systematic biases of the result.
Actually, the correction of systematic effects can generally be obtained without performing on data
the maximal transfer of events (predicted by the transfer matrix), but only part of it.
The method described in this paper allows one to control the amount of these correlations
and an optimization can be done with respect to the potential systematic bias of the final result.

This method is to be applied on binned, one dimensional data and the label of the physical 
quantities will be provided by the bin numbers.
It can be directly generalized to multidimensional problems.
For the sake of simplicity, we will consider only examples where the number of bins of the
data and reconstructed MC is the same as the one for the unfolded distribution and true MC.
However, it is rather easy to generalize the method to include a binning change in the 
unfolding operation.
All the formulae in the following will be given for this general case.
We suppose the statistics of the data and of the simulation high enough to allow a 
``reasonable'' separation into a series of bins 
% (i.e. except for the regions where the true and/or reconstructed spectra are at a relatively low level, 
% the other bins should be filled)
.
If this is not the case, the size of the bins must be adapted to the available statistics.

\section{The regularization function}
\label{Sec:RegF}
In order to dynamically reduce fluctuations and not to unfold events which could be due to
fluctuations, in particular from the subtracted background, one can use a regularization function $f(\Delta x,\sigma,\lambda)$.
This function provides an information on the significance of the absolute deviation $\Delta x$ between
data and simulation in a given bin, with respect to the corresponding error $\sigma$.
It must be a smooth monotonous function going from 0, when $\Delta x = 0$, to 1, when
$\Delta x >> \sigma$.
$\lambda$ is a scaling factor, used as a regularization parameter. 
As we will see in the following, changing the regularization function that is used in our method will change the 
way we discriminate between real deviations and statistical fluctuations.
It can be seen as the complementary of a generalized (by the use of $\lambda$) 
``p-value'' of a given deviation between two variables affected by errors, if one considers 
the hypothesis that the MC describes well the data in the given bin.
Actually, for $\lambda = 1$ this function is the complementary of the p-value defined 
in \cite{Amsler:2008}, computed for a given PDF.
If the correlations between the errors of the values entering the computation of $\Delta x$
are not negligible, only the uncorrelated part must be kept when computing $\sigma$.

\begin{figure}[h]
\begin{center}
\includegraphics[width=14cm]{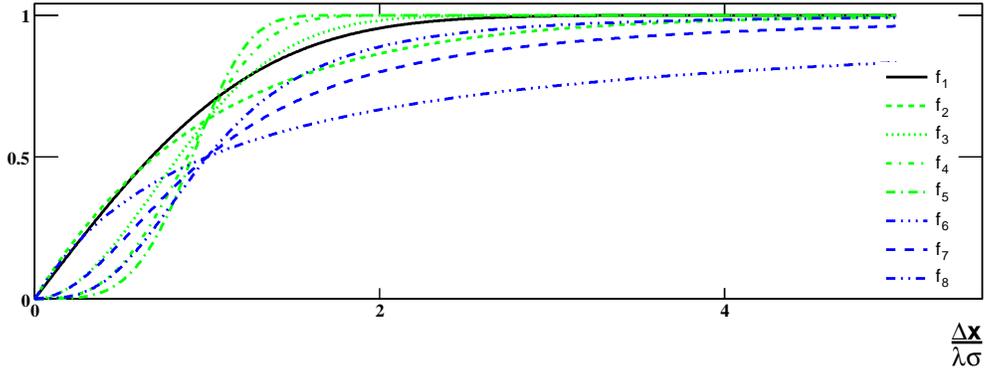}
\caption{Behaviour of the functions $f_{1..8}$ with respect to ${\Delta x}/{(\lambda \sigma )}$.}
\label{fig:functions}
\end{center}
\end{figure}

For the unfolding procedure, we consider several types of functions of the relevant variable ${\Delta x}/{(\lambda \sigma )}$ :
\begin{eqnarray}
\label{f:regFunctions}
&& f_1 = Erf \left( \frac{\Delta x}{\sqrt{2} \lambda \sigma} \right), \\
\label{f:1Mexp}
&& f_{1+n} = 1 - e^{- \left( \frac{\Delta x}{\lambda \sigma } \right) ^n }, 
\textrm{with }n \in \{ 1;2;3;4 \}, \\
\label{f:1Mfrac}
&& f_{5+n} = 1 - \frac{1}{ 1 + \left( \frac{\Delta x}{\lambda \sigma } \right) ^n }, 
\textrm{with }n \in \{ 1;2;3 \}.
\end{eqnarray}
Their dependence on the relevant ratio is shown in Fig. \ref{fig:functions} .
In general, we will use different $\lambda$ parameters for the regularization function for each step
of the unfolding procedure.
We will see however that some of these parameters can be unified or even dropped.

\section{Monte Carlo normalization procedure}
\label{Sec:MCnorm}
In the unfolding procedure we will need to perform a comparison of data and reconstructed 
MC, as well as true MC and intermediate unfolded spectra.
We must not forget that the data may contain structures, which were not (well) simulated 
in the MC.
If the number of events corresponding to them is not negligible with respect
to the total number of events in the data, the normalization of the MC spectra should be
determined with care.
One must actually ignore the ``unknown'' (to the simulation) events in these structures when computing the 
normalization factor.
Not doing so (a standard normalization procedure would use the total number of events in the two spectra) 
could result in generating fake differences between the two spectra, in regions where the data are well 
described by the simulation.

For a good normalization, one needs to get an estimation of the number of events in the 
data/intermediate unfolding spectrum which correspond to features that were simulated 
in the reconstructed/true MC ($NE_{dSmc}$).
Dividing $NE_{dSmc}$ by the number of events in the true and reconstructed MC spectra ($NE_{mc}$)
one gets the normalization factor to be used for the MC.
This is done as follows: \\
1) A first estimation of this number is provided by the total number of events in the
data and in the intermediate unfolding distribution ($NE_d$), minus those which are potentially due 
to background subtraction:
$NE_{dSmc} = \sum_{k=1}^{NB_{d}}{\left( d_k-bd_k \right)} = \sum_{k=1}^{NB_{u}}{\left( u_k-bu_k \right)}$. 
Here $NB_{d}$ is the number of bins in the data and in the reconstructed MC distribution, $NB_{u}$ is 
the number of bins in the unfolding result and in the true MC distribution, $d_k$ stands for the number of 
reconstructed data events in the bin k (after background subtraction) and $bd_k$ for the number of events 
in the bin k of the data distribution which are potentially due to a fluctuation in the subtracted background,
$u_j$ stands for the number of unfolded events in the bin j and
$bu_j$ for the number of events in the bin j of the unfolded distribution which are potentially due
to a fluctuation in the subtracted background.\\
One can then improve this estimation by an iterative procedure: \\
2) A better estimation is provided by
\begin{equation}
(NE_{dSmc})' = NE_{dSmc} + \sum_{k=1}^{NB_{d/u}}{
( 1 - f \left( \left| \Delta (d/u)_k \right| , \tilde{\sigma} (d/u)_k , \lambda_N \right) ) 
\cdot \Delta (d/u)_k} ,
\label{eq:NEprdSmc}
\end{equation} 
with $\Delta (d/u)_k = (d/u)_k - (bd/bu)_k - \frac{NE_{dSmc}}{NE_{mc}} \cdot (r/t)_{MCk}$ where
$r_{MCk}$ is the number of reconstructed MC events and $t_{MCk}$ the number of true MC events, and 
$\tilde{\sigma} (d/u)_k = \sqrt{\sigma^2((d/u)_k) + (\frac{NE_{dSmc}}{NE_{mc}})^2\cdot\sigma^2((r/t)_{MCk})}$ 
if the errors of $d_k$ or $u_k$ ($\sigma((d/u)_k)$) are not correlated with the errors of $r_{MCk}$ 
or respectively $t_{MCk}$ ($\sigma((r/t)_{MCk})$).
This hypothesis is true for the first-step comparison of data and reconstructed MC.
The situation is different at further iteration steps and for the comparison of true MC and 
intermediate unfolding spectra.
There, only their uncorrelated parts must be added quadratically to get $\tilde{\sigma}_k$. 
In practice, however, as the MC errors are generally negligible, considering only the 
data/unfolding errors at this level may be sufficient. \\
Using Eq. (\ref{eq:NEprdSmc}), one adds to the previous estimate other events
which have a low probability to belong to initially unknown structures. 
On the contrary, due to the factor in front of $\Delta (d/u)_k$, 
there is only a relatively small contribution from events likely to belong to initially unknown structures. \\
In the special cases when $(d/u)_k - (bd/bu)_k < 0$ or $(r/t)_{MCk} = 0$, the content of
the corresponding bins will be neglected when computing the number of events for the data
and MC. 
This kind of situation will also receive a special treatment in the unfolding procedure,
the estimation of the background fluctuations and the improvement of the transfer matrix. \\
3) Replace $NE_{dSmc}$ by $(NE_{dSmc})'$ and iterate until the change in $NE_{dSmc}$ between two 
consecutive iterations gets very small. \\
In the examples presented in this paper we stopped the normalization iterations when the relative
improvement of two consecutive steps was less than $10^{-6}$.

It should be emphasized that a better estimation of $NE_{dSmc}$ improves the separation 
between ``known'' and ``unknown'' structures, which then improves the next estimation.
At the same time, the values of the parameter $\lambda_N$ (for normalization), providing a good convergence 
of the normalization procedure, need to be studied, as described in Section \ref{Sec:OptParam}.
The convergence properties of the normalization procedure also depends on the amount
of background fluctuations which are not yet identified when the normalization is performed.
In the normalization procedure, these fluctuations, corresponding to data points with larger 
error bars, and would more likely not be identified as new structures.
As the estimation of the remaining background fluctuations gets improved, the normalization
follows.

At this level we made a distinction between $bd_k$ and $bu_k$ only
to keep the method as general as possible.
Indeed, if the binning of the initial (data) and final (unfolded) distributions are not the same,
one must build a prescription to convert one vector into another, by a rebinning (in any case, no 
unfolding will be done for these events).
Using a rebinning transformation $R$ to pass from data to unfolding bins and the
inverse rebinning transformation $R'$ to pass from unfolding to data bins, 
one can convert one vector into the other, through:
\begin{eqnarray}
\label{eq:bu_j}
bu \xrightarrow{\text{   }R\text{   }} bd , \\
\label{eq:bd_i}
bd \xrightarrow{\text{   }R'\text{  }} bu .
\end{eqnarray}
These two transformations, if valid for all the sets of possible spectra in the two binnings 
(as defined before), must be non-linear.
Actually, if they were linear, as their two matrix would contain only positive elements, 
even if $NB_d = NB_u$, their product could not be equal to the identity matrix.
If the binning of the data and unfolding result are the same, the two vectors will always
be identical (i.e. $R = R' = I$).

\section{Folding and unfolding}
\label{Sec:FoldAndUnf}
In the MC simulation of the detector one can directly determine the number of events 
which were generated in the bin $j$ and reconstructed in the bin $i$ ($A_{ij}$).
Provided that the transfer matrix $A$ gives a good description of the detector effects, it is 
straightforward to compute the corresponding folding and unfolding matrix:
\begin{eqnarray}
\label{eq:P}
P_{ij}  &=& \frac{A_{ij}}{\sum_{k=1}^{NB_{d}}{A_{kj}} } , \\
\label{eq:Pp}
\tilde{P}_{ij} &=& \frac{A_{ij}}{\sum_{k=1}^{NB_{u}}{A_{ik}} } .
\end{eqnarray}
The folding probability matrix, as estimated from the MC simulation, $P_{ij}$ gives the probability 
for an event generated in the bin $j$ to be reconstructed in the bin $i$.
The unfolding probability matrix $\tilde{P}_{ij}$ corresponds to the probability for the ``source'' 
of an event reconstructed in the bin $i$ to be situated in the bin $j$.
These matrix elements are well defined if the true MC bin j, and respectively the reconstructed MC 
bin i, are not empty.
If that is not the case, one can set these values to zero and a special treatment must be 
provided for these bins in the unfolding algorithm.

The folding matrix describes the detector effects, and one can only rely on the simulation
in order to compute it.
The quality of this simulation must be the subject of dedicated studies within the analysis.
As a result of these studies, the transfer matrix can be improved, a systematic error is to be
associated to it and must be propagated to the unfolding result.
The unfolding matrix depends not only on the description of detector effects but also 
on the quality of the model which was used for the true MC distribution.
It is actually this model which can (and will) be iteratively improved, using the comparison 
of the true MC and unfolded distributions.

It is easy to see that the reconstructed and true MC distributions are related through:
\begin{eqnarray}
\label{eq:rMC}
r_{MCi} &=& \sum_{k=1}^{NB_{u}}{P_{ik} \cdot t_{MCk}} , \\
\label{eq:tMC}
t_{MCj} &=& \sum_{k=1}^{NB_{d}}{\tilde{P}_{kj} \cdot r_{MCk}} .
\end{eqnarray}
Provided that the detector effects are well simulated, the folding equation (\ref{eq:rMC}) will
also stand for data, no matter what the true spectrum is in the real world.
The unfolding equation (\ref{eq:tMC}) can be written for data only in an approximate way,
due to the differences between the model used for $t_{MC}$ and the real physical signal
which introduce an uncertainty on the initial unfolding matrix.
However, $t_{MC}$ and hence the unfolding matrix, can be iteratively improved and therefore it is an improved 
version of (\ref{eq:tMC}) that will be used to compute the unfolding result from data.
Other effects will also be taken into account in the unfolding equation, as we shall see in the next
subsection.

\subsection{The general one-step unfolding procedure}
\label{sec:unfold}
Equation (\ref{eq:tMC}) can be seen as a ``basis for the unfolding'', in the sense that it
describes this operation performed on the MC.
If the differences between data and reconstructed MC are relatively small, 
then those between the unfolding result and true MC will be small too.
In any case, one can start with an unfolded distribution equal to the true MC, and add to it
the unfolded data events which are not present in the reconstructed MC.
For the initial true MC one can use either a previously measured spectrum (if available) or
one inspired by the data and the transfer matrix.

In order to perform the unfolding, one must first use the iterative procedure described in 
Section \ref{Sec:MCnorm} to determine the MC normalization coefficient (${NE_{dSmc}}/{NE_{mc}}$).
At this level, it is the comparison of the reconstructed MC with the data that is used in the 
normalization procedure.

One can then proceed to the unfolding, where, in the case of identical initial and final binnings, 
the result for $j\in [1;NB_u]$ is given by:
\begin{eqnarray}
 u_j &=& t_{MCj}\cdot \frac{NE_{dSmc}}{NE_{mc}} + bu_j \nonumber \\
  && + \sum_{k=1}^{NB_{d}}{ 
  \begin{cases}
    f \left( \left| \Delta d_k \right| , \tilde{\sigma} d_k , \lambda\right)
    \cdot \Delta d_k \cdot \tilde{P}_{kj} +
    \left( 1 - f \left( \left| \Delta d_k \right| , \tilde{\sigma} d_k , \lambda\right) \right)
    \cdot \Delta d_k \cdot \delta_{kj} \textrm{, if $r_{MCk}\ne 0$} \\ 
				  \hspace{9,2cm}
    				  \textrm{and $d_k - bd_k > 0$;} \\
    \Delta d_k \cdot \delta_{kj} \textrm{, if $r_{MCk} = 0$ or $d_k - bd_k \le 0$,}
  \end{cases}
    }
\end{eqnarray}
where the same notation as in (\ref{eq:NEprdSmc}) is used.
In the case of different binnings for the data and the unfolding, the Kroneker symbol $\delta$ 
must be replaced by a rebinning transformation $R$.

The first two contributions to the unfolded spectrum are given by the normalized true MC and
the events potentially due to a fluctuating background subtraction, which are not transferred 
from one bin to another.
Then one adds the events which are present in the data minus the potential effect from 
the background, and which are absent in the normalized reconstructed MC.
A fraction $f$ of these events are unfolded using the estimate of the unfolding probability 
matrix $\tilde{P}$, and for the rest, only a rebinning is done (if necessary).
With the description of the regularization functions given in Section \ref{Sec:RegF}, it is 
clear that reducing $\lambda$ would result in increasing the fraction of unfolded events,
and reducing the fraction for which only a rebinning is done.
This unfolding step benefits from two regularizations: one through the use of the previously 
described functions, and a second (implicit) one, through the mixing of events in different bins,
by transfer of events.
Changing $\lambda$ can reduce one regularization and enhance the other.
Choosing an appropriate value for this coefficient provides one with a dynamical attenuation of
spurious fluctuations, without reducing the efficiency of the unfolding itself.
This is possible due to the fact that the fraction of unfolded events is larger when 
the deviations between the spectra are larger with respect to the corresponding errors.

Some precautions need to be taken if $d_k - bd_k \leq 0 $ or $r_{MCk} = 0$.
The first situation can typically occur if the background subtraction that was made for data, 
in the given bin, is too large, or if the binning is not well adapted to the data statistics.
The second one indicates that the given data bin size is too small even for the MC statistics and 
the corresponding unfolding elements can not be computed.
In these two situations, the events which are present in the data minus the potential remaining 
background and minus the normalized reconstructed MC will be simply rebinned 
(or kept in their original bin if the data and unfolding binnings are identical).

\subsection{The estimation of the remaining background fluctuations}
\label{Sec:estBkg}
After having performed one unfolding procedure yielding to a result $u$, as described before, 
one can directly compare it with the true MC.
Doing this, one can compute an estimate of the fluctuations from subtracted background events.
As these fluctuations must be identical in the reconstructed data and in the unfolding result,
it is important that the first unfolding does not affect them (hence a constraint on the 
corresponding parameter $\lambda_S$ (for subtraction)).

Just as for the unfolding, one must first use the iterative procedure described in Section 
\ref{Sec:MCnorm} to determine the MC normalization coefficient (${NE_{dSmc}}/{NE_{mc}}$), 
looking this time for structures present in the unfolding result, but not simulated in the true MC.

Then one can (re-)estimate the background to be subtracted
\begin{equation}
\label{eq:imprbu}
bu'_j = \left( 1 - f \left( \left| \Delta' u_j \right| , \tilde{\sigma} u_j , \lambda_S\right) \right) 
\cdot\Delta' u_j ,
\end{equation}
where $\Delta' u_k = u_k - \frac{NE_{dSmc}}{NE_{mc}} \cdot t_{MCk}$.

Here, increasing $\lambda_S$ would increase the fraction of not-simulated events, which will 
be assigned to the estimate of the remaining background from fluctuations in the subtraction.
In principle one could perform this estimation using $\Delta$, with the right side of (\ref{eq:imprbu}) increased
by the previous estimate ($bu_j$), but here we rather use $\Delta'$ so that we avoid an artificial increase 
of $bu'$ in regions where the spectra differ only due to a normalization effect.
This allows one to use sufficiently large values of $\lambda_S$, to efficiently subtract 
real remaining background fluctuations.
The use of large values of $\lambda_S$ will also prevent the propagation and potential 
amplification of the fluctuations in the unfolding result.

Once this estimate is obtained, one can use it to get a better normalization and then reestimate
the background fluctuations.
This new information is however exploited very fast and one or two iterations are largely enough at
this level.

One can perform the (re-)estimation (\ref{eq:imprbu}) only in the usual case, when
$u_j - bu_j > 0$ and $t_{MCj} \ne 0$.
Otherwise, the estimation of the remaining background is trivially modified
\begin{eqnarray}
\label{eq:imprbuPath}
bu'_j = \Delta' u_j \nonumber ,
\end{eqnarray}
implying a fixation of the corresponding ``pathological'' events at the next unfolding step.
This solution seems natural if the unfolding result minus the remaining background has a negative
number of events in the given bin.
Such a situation can occur due to an overestimation of the background that was subtracted to get
the data spectrum to be unfolded.
If the origin of the problem is the low MC statistics in the neighbourhood of the given bin,
a more suitable solution could be the choice of an adapted binning.
We feel that this problem should be dealt with in unfolding methods.

\subsection{The improvement of the unfolding probability matrix}
\label{sec:impMatrix}
As explained in the introduction, if the initial true MC distribution does not contain or badly describes
some structures which are present in the data, one can iteratively improve it, and hence 
the transfer matrix.
This can be done by using a better (weighted) true MC distribution, with the same folding matrix
describing the physics of the detector, which will yield to an improved unfolding matrix.

Just as in the previous section, one must determine the MC normalization coefficient, looking for 
structures present in the unfolding result, but not simulated in the true MC.
As the description of the new structures by the MC will get improved, the task of the normalization 
procedure will be easier.

Then the improvement is performed for one bin $j$ at the time. 
If however $u_j - bu_j < 0$ or $t_{MCj} = 0$, one does not modify the transfer matrix in 
the corresponding true bin.
Generally, this is not the case, and one can get an improved determination of all the elements 
of the column $j$ of the transfer matrix
\begin{equation}
\label{eq:imprA}
A'_{ij} = A_{ij} + f \left( \left| \Delta u_j \right| , \tilde{\sigma} u_j , \lambda_M\right) \cdot\Delta u_j 
\cdot \frac{NE_{mc}}{NE_{dSmc}} \cdot P_{ij} \textrm{, for $i\in \{ 1; NE_d\}$ .}
\end{equation}
Here, $\lambda_M$ (for modification) stands for the regularization parameter used when modifying the matrix.
Increasing $\lambda_M$ would reduce the fraction of events in $\Delta u_j$ used to improve
the transfer matrix.

This method allows an efficient improvement of the folding matrix, without introducing spurious 
fluctuations.
This is due to the fact that the larger the difference between the spectra, with respect to the 
corresponding errors, the larger the fraction of events used for the improvement 
of the matrix will be.
The amplification of small fluctuations can be prevented at this step of the procedure too.

\section{The iterative unfolding strategy}
\label{Sec:IterUnfStrategy}
In this section we describe a quite general unfolding strategy, based on the elements presented before.
It works for situations presenting all the difficulties listed before, even in a simplified form,
where some parameters are dropped and the corresponding steps get trivial.
The strategy can be simplified even more, for less complex problems.

One will start with a null estimate of the background.
A first unfolding, as described in Section \ref{sec:unfold}, is performed, 
with a relatively large value of $\lambda = \lambda_L$.
This step will not produce any important transfer of events from the regions with potential 
remaining background (provided that $\lambda_L$ is large enough).
Actually, it is the condition that this transfer must be very small that will impose a minimal 
value for $\lambda_L$.
The first unfolding will perform the correction equal to the difference between the normalized 
true and reconstructed MC.
This is generally the main correction that the unfolding has to provide.
Concerning the new structures in the data, depending on their relative size, on the size of 
the transfer corrections and on $\lambda_L$, the corresponding corrections at this level can be 
more or less important.

One can even drop $\lambda_L$, by taking its value very large, which amounts to performing 
only the main correction discussed in the previous paragraph.
This obviously guarantees the stability of the remaining background fluctuations for this step.

At this level one can start the iterations: \\
1) Estimation of the remaining background fluctuations \\
An estimate of the remaining background fluctuations can be obtained using the procedure
described in Section \ref{Sec:estBkg}.
The parameter $\lambda_S$ used here must be large enough, not to underestimate the remaining 
background.
As explained before, this can also prevent unfolding fluctuations from propagating. 
$\lambda_S$ can however not be arbitrary large, as this operation must not bias initially
unknown structures, by not allowing their unfolding.
The ability of the method to simultaneously satisfy to these conditions clearly depends on the
properties of the function $f_n$, as it defines the degree of significance of deviations
between spectra. \\
2) Improvement of the unfolding probability matrix \\
Using the method described in Section \ref{sec:impMatrix}, one can improve the folding
matrix $A$ and get a better estimate of the remaining background.
A parameter $\lambda_M$, small enough for an efficient improvement of the matrix,
yet large enough not to propagate spurious fluctuations (if not eliminated at another step), 
must be used at this step. \\
3) An improved unfolding \\
A parameter $\lambda = \lambda_U$ will then be used to perform an unfolding following Section 
\ref{sec:unfold}, exploiting the improvements done at the previous step. 
It must be small enough to provide an efficient unfolding, but yet large enough to avoid
spurious fluctuations (if not eliminated elsewhere).\\
These three steps will be repeated until one gets a good agreement between data and reconstructed MC
plus the estimate of the background.
Another way of proceeding could consist in stopping the iterations when the improvement brought by the last one
on the intermediate result is relatively small.
If the parameters of the unfolding are well chosen, the two conditions yield similar results.
Provided that one has a good model of the true data, one can even use a toy simulation to estimate 
the needed number of steps.

The values of the $\lambda$ parameters are to be obtained from toy simulations, as will be 
shown in the following.
As one does not have a prior lower limit for $\lambda_M$ and $\lambda_U$, other than the one
related to the regularization, one can discard them by taking their values to zero 
(i.e. using all the data - reconstructed MC difference in the unfolding, and all the difference 
between the true MC and the intermediate unfolding, to improve the transfer matrix).
This can be done if the higher limit on $\lambda_S$ provides enough regularization at the 
level of the subtraction procedure.

In principle, one could also perform an estimation of the remaining background fluctuations,
when comparing the data with the reconstructed MC.
This could be done as a straightforward adaptation of the method explained in Section \ref{Sec:estBkg} 
or computing this estimate as an improvement of the previous one, with $\Delta' d$ replaced by
$\Delta d$.
This second method could be reliable if the normalization effect is small enough.
However, when doing the estimation of the background fluctuations at the data - reconstructed MC level, 
one would not benefit from the iterative improvement of the true MC.
This can be useful in the method of Section \ref{Sec:estBkg}, as it improves the separation between 
real new structures and fluctuations, especially when the resolution effects are important.

\section{A complex example for the use of the unfolding procedure}
\label{Sec:ComplexEx}
In the following we describe a rather complicated, yet realistic test,
proving the robustness of the method.
It exhibits all the features discussed previously, which are simultaneously taken into account
by the unfolding.
For the clarity of the presentation, the structures and dips of the spectrum are however separated.
The true and reconstructed spectra for this test have 200 identical bins.

\subsection{Building the data spectra and transfer matrix}
\label{subsec:building1}
\begin{figure}[h]
\begin{center}
\includegraphics[width=10cm]{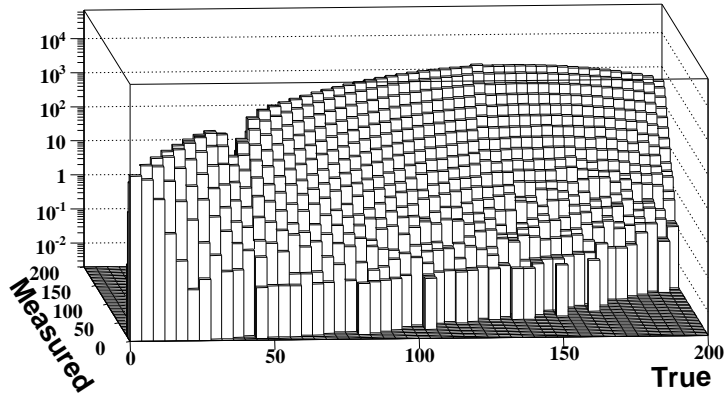}
\caption{Model $\bar{A}$ of transfer matrix. Here, for display reasons, the bins are four times larger 
than the real ones (an average was computed inside squares of $4X4$ initial bins).}
\label{fig:Ax4}
\end{center}
\end{figure}

The first step consists in building a model for the transfer matrix $\bar{A}$.
The transfer matrix 
\begin{eqnarray}
\label{eq:Tmatrix}
&& T_{ij} = \frac{N_T}{2\pi \sigma_x \sigma_y}
	 \left( 1 + 0.2 e^{-\frac{(j-130)^2}{2 \sigma_r^2}} - e^{-\frac{(j-40)^2}{2 \cdot \sigma_d^2}} \right)
	 e^{-\frac{(j-100)^2}{2 \sigma_x^2} -\frac{(i-j)^2}{2 \sigma_y^2} } ,
\end{eqnarray}
with $N_T = 10^6$, $\sigma_x = 40$, $\sigma_y = 3, \sigma_r = 1$ and $\sigma_d = 4$;
exhibits resolution effects and corresponds to a true spectrum which is mainly Gaussian. 
In addition, this spectrum includes a Gaussian resonance (more narrow than resolution) 
centred on the bin 130 and a dip around the bin 40 (larger than resolution).
One then includes a systematic transfer of events from high to lower bins, getting the final model
matrix (see Fig. \ref{fig:Ax4}):
\begin{equation}
\label{eq:Amatrix}
\bar{A}_{kj} = \sum_{i=k}^{N}{T_{ij}\cdot c_1 \cdot \left( \frac{k+1}{i+1} \right)^4 \cdot e^{-(i-k)/c_2} } ,
\end{equation}
with $c_1 = 2$ and $c_2 = 8$.
In practice, one can find an example of this type of transfer of events, for measurements of the cross section 
$e^+e^- \to hadrons(n\gamma)$ with the radiative return method.
Indeed, not detecting a final state photon will result in an underestimation of the mass of the 
given event. 

It is a fluctuation of the model that will be used as MC transfer matrix $A$ for the unfolding.
The model itself provides the ``real'' detector matrix which will be crucial for the building and 
folding of the true distribution.

One must then fix a normalisation ratio between data and MC simulation (of the order of 1/10
for usual experiments).
The true data distribution will be given by the normalised true distribution corresponding to
$\bar{A}$, completed with structures unknown by the MC (a ``bias''):
\begin{equation}
bias_j = \frac{2\cdot 10^4}{3\cdot \sqrt{2\pi}} \cdot
	\left[ e^{-\frac{(j-90)^2}{2\cdot \sigma_1^2} } 
	+ e^{-\frac{(j-170)^2}{2\cdot \sigma_2^2} } \right] ,
\label{eq:bias1}
\end{equation}
with $\sigma_1 = \sqrt{6}$ and $\sigma_2 = 4$.
The resonance centred on the bin 90 is more narrow than resolution and will therefore be hard
to correct, whereas the one centred in 170 is larger.

The folding probability matrix is to be computed from $\bar{A}$.
At this point, one can compute the known data ($dk$), by folding the true data without the new structures.
The folding of the true data yields to the reconstructed spectrum (after fluctuation within 
statistical errors).
The final data spectrum (filled circles in Fig. \ref{fig:Unf1}) is obtained after adding a remaining background component, 
where it can typically appear in practice, i.e. in the dip of the spectrum, with the average per bin:
\begin{equation}
\label{eq:bkgd}
bd_i =  \frac{7\cdot 10^3}{3\cdot \sqrt{2\pi}} \cdot e^{-\frac{(i-40)^2}{2\cdot \sigma_b^2} } ,
\end{equation}
with $\sigma_b = \sigma_d$.
For the background, we attribute an error bar equal to its average value in each bin.
This background can come with or without statistical fluctuations corresponding to its error bars.
In order to simulate this feature, we fluctuate the background within its errors for $i \leq 40$ only,
and we add it to the data fluctuated within their statistical errors.
The final errors of the data, which will enter the unfolding algorithm, are given by the 
quadratic sum of the errors of the background plus the statistical ones from the data without 
background.

The unfolding result is to be compared with the true data plus the remaining background.
This represents a test over a very large dynamic scale.
The results that will be presented for this test were obtained using the function (\ref{f:1Mexp}), for
$n = 2$, but similar results were obtained with the other functions.

\subsection{Optimization of the parameters}
\label{Sec:OptParam}

The parameters to be used in the unfolding were studied with an independent toy simulation, with 
various parameters for the fluctuations in background subtraction, and the bias.
In practice, one would need approximate values of the parameters for the spectrum and the potentially 
dangerous background fluctuations, in order to perform the further study, to determine the parameters of the unfolding.
One can get a very preliminary estimate for the parametrization of the spectrum from the difference between data 
and reconstructed MC, together with the transfer matrix, and from the procedure of background subtraction.
Once the unfolding performed, the determination of the spectrum parameters is also drastically improved.
In principle, at that point, one could try to re-estimate the optimal parameters of the unfolding and
try to improve them with more accurate toy simulations.
In practice however, excepting very extreme situations, this new estimate 
will be consistent with the first one.

\begin{figure}[h]
\begin{center}
\includegraphics[width=5cm]{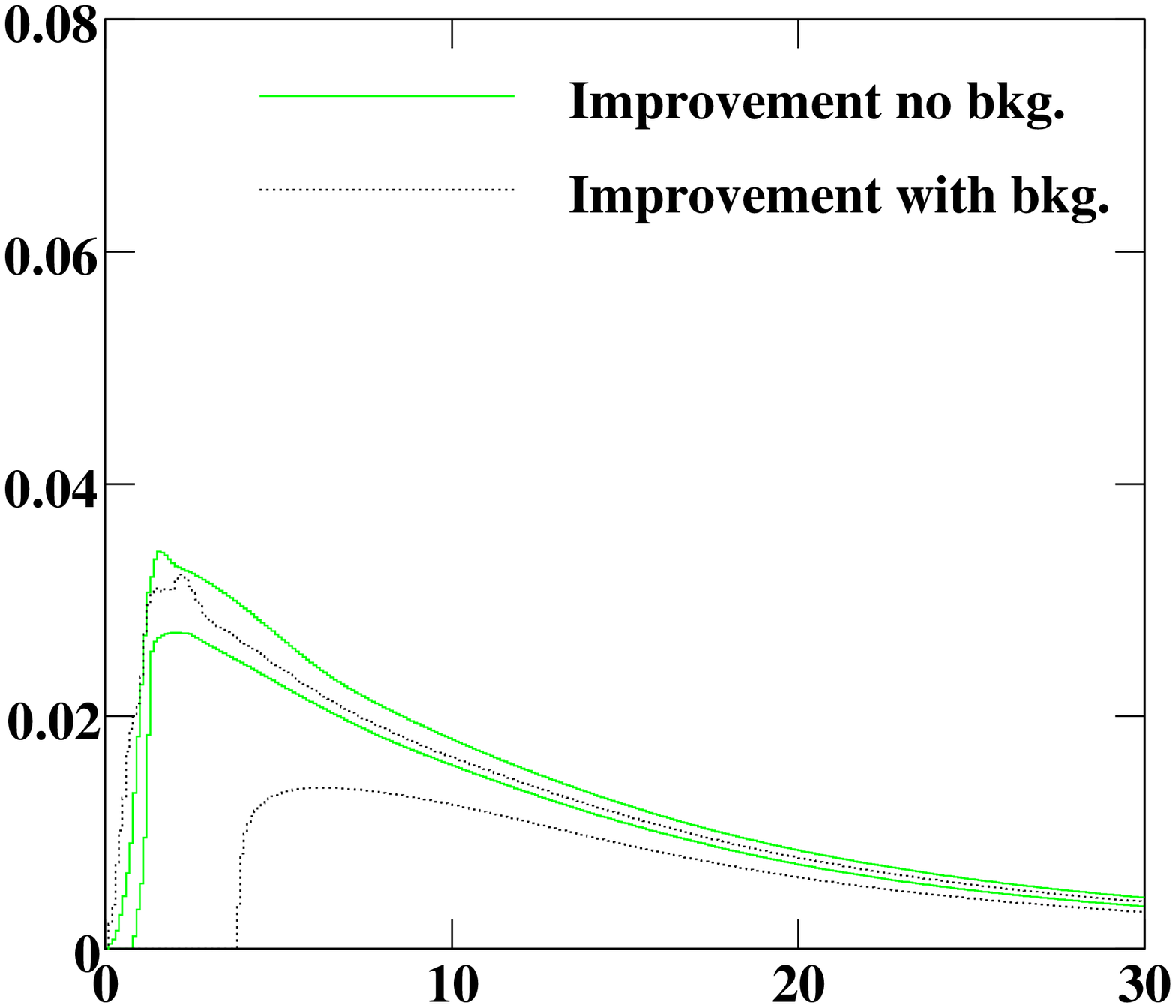}
\includegraphics[width=5cm]{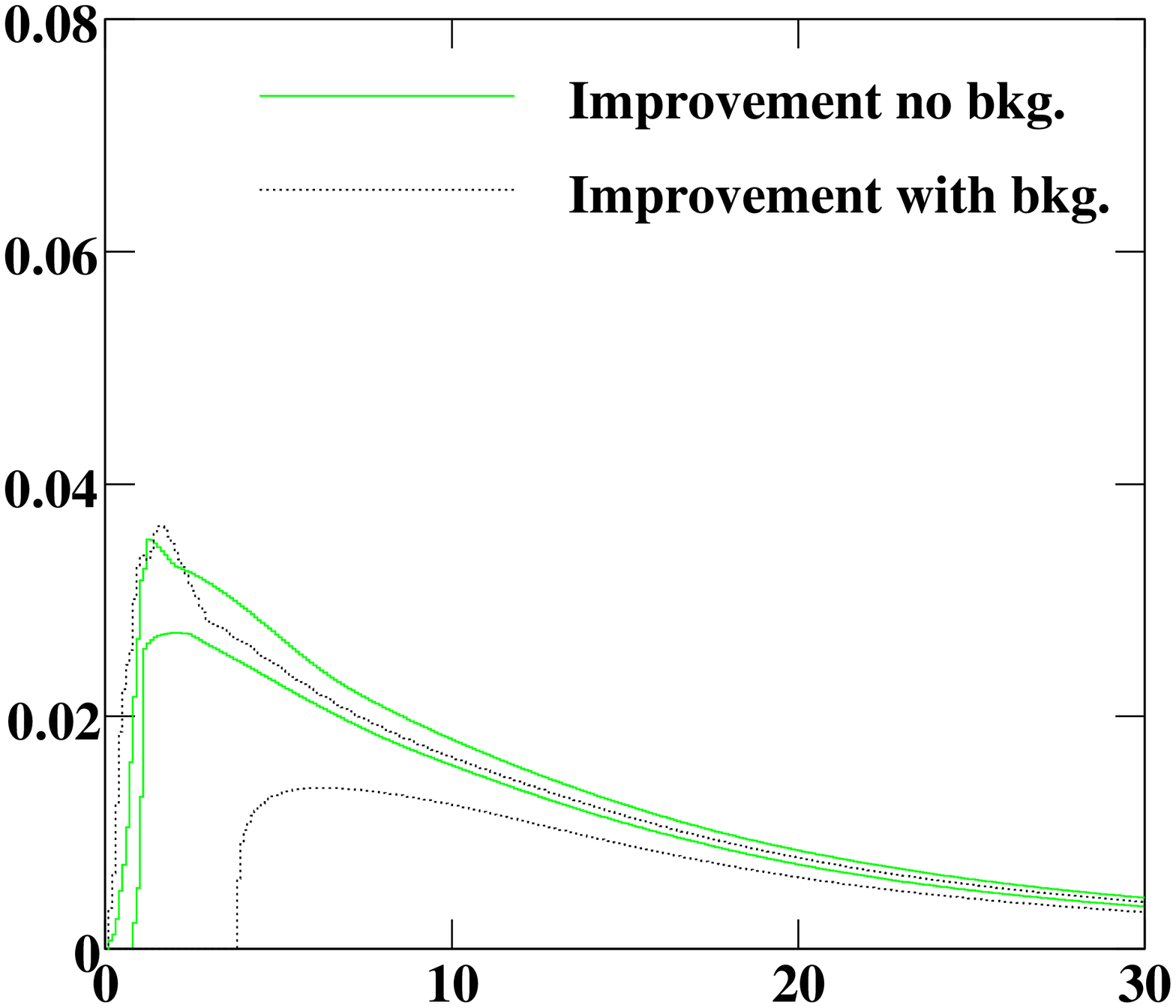}
\includegraphics[width=5cm]{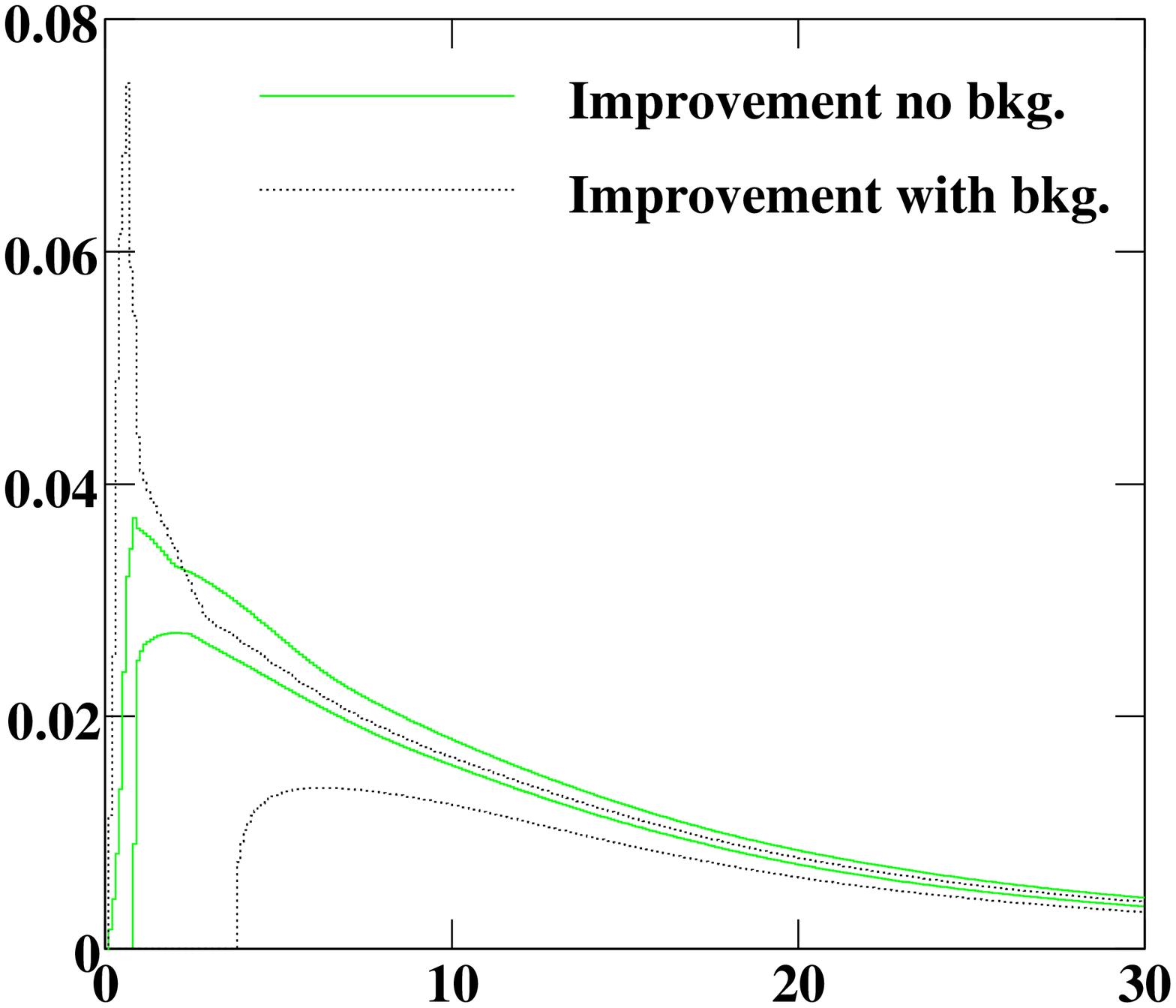} 
\caption{Normalization relative improvement limits as a function of $\lambda_N$, without background or with 
the usual background value, obtained after, at most, from left to right, 30, 50 and respectively 200 steps.}
\label{fig:optLnUB}
\end{center}
\end{figure}
\begin{figure}[h]
\begin{center}
\includegraphics[width=5cm]{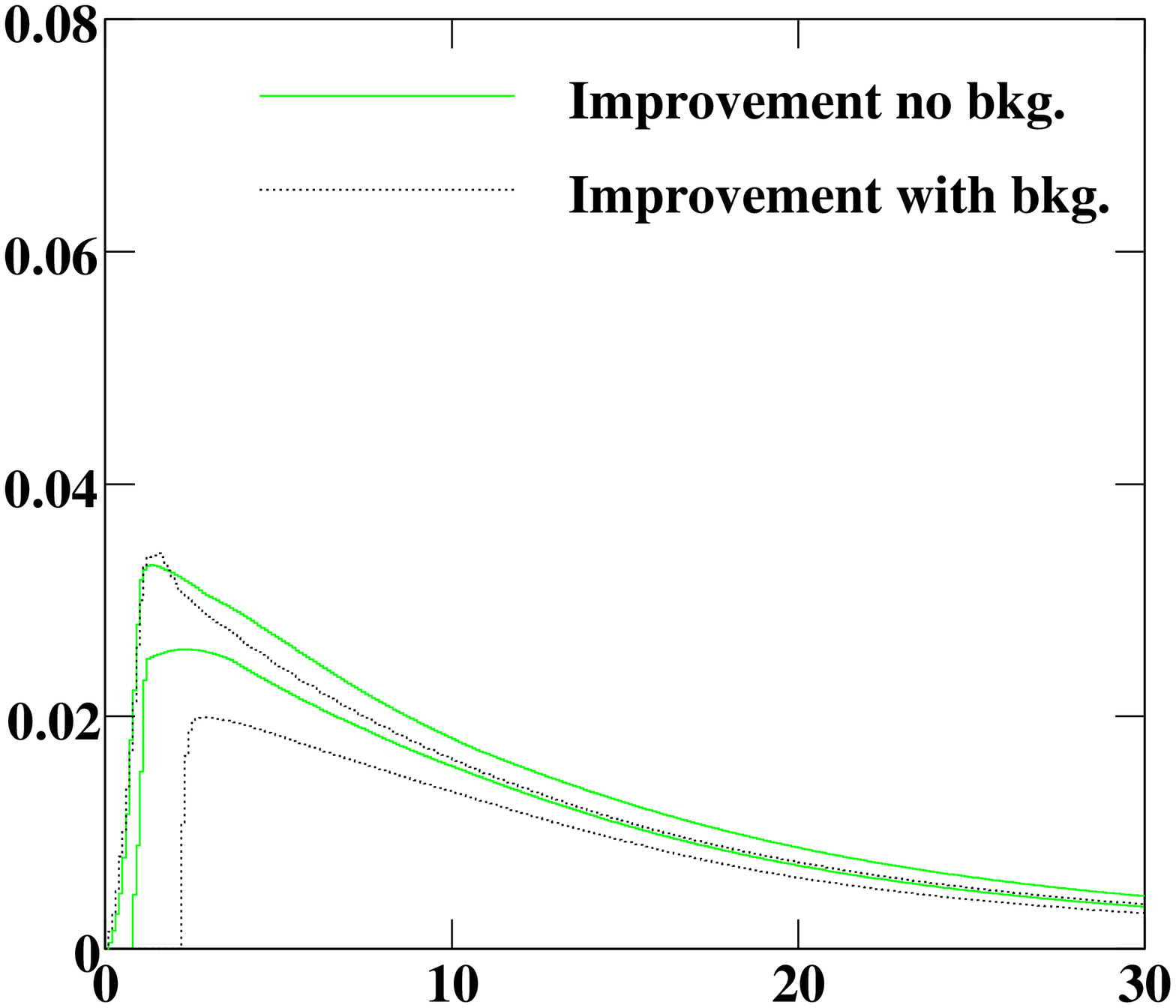}
\includegraphics[width=5cm]{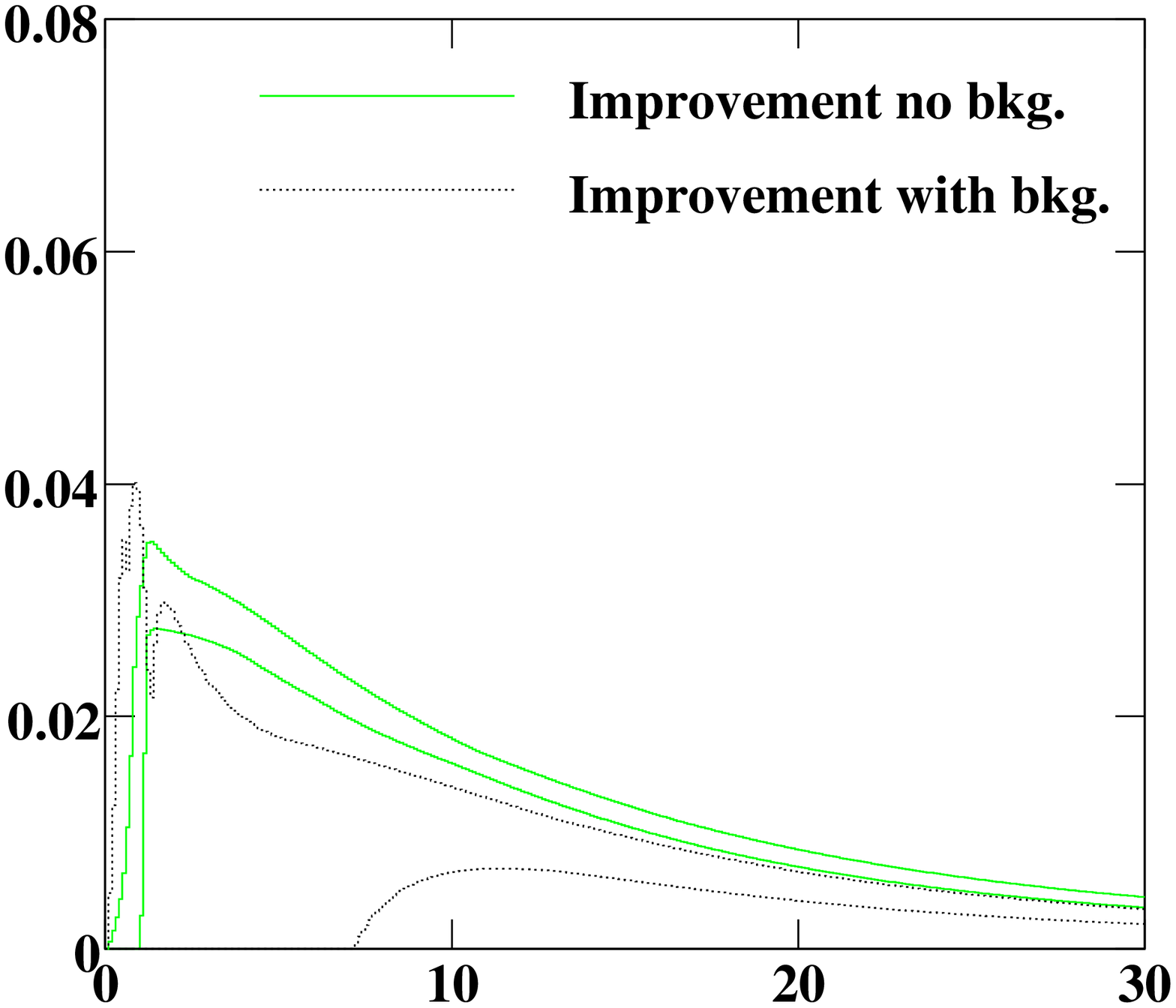}
\includegraphics[width=5cm]{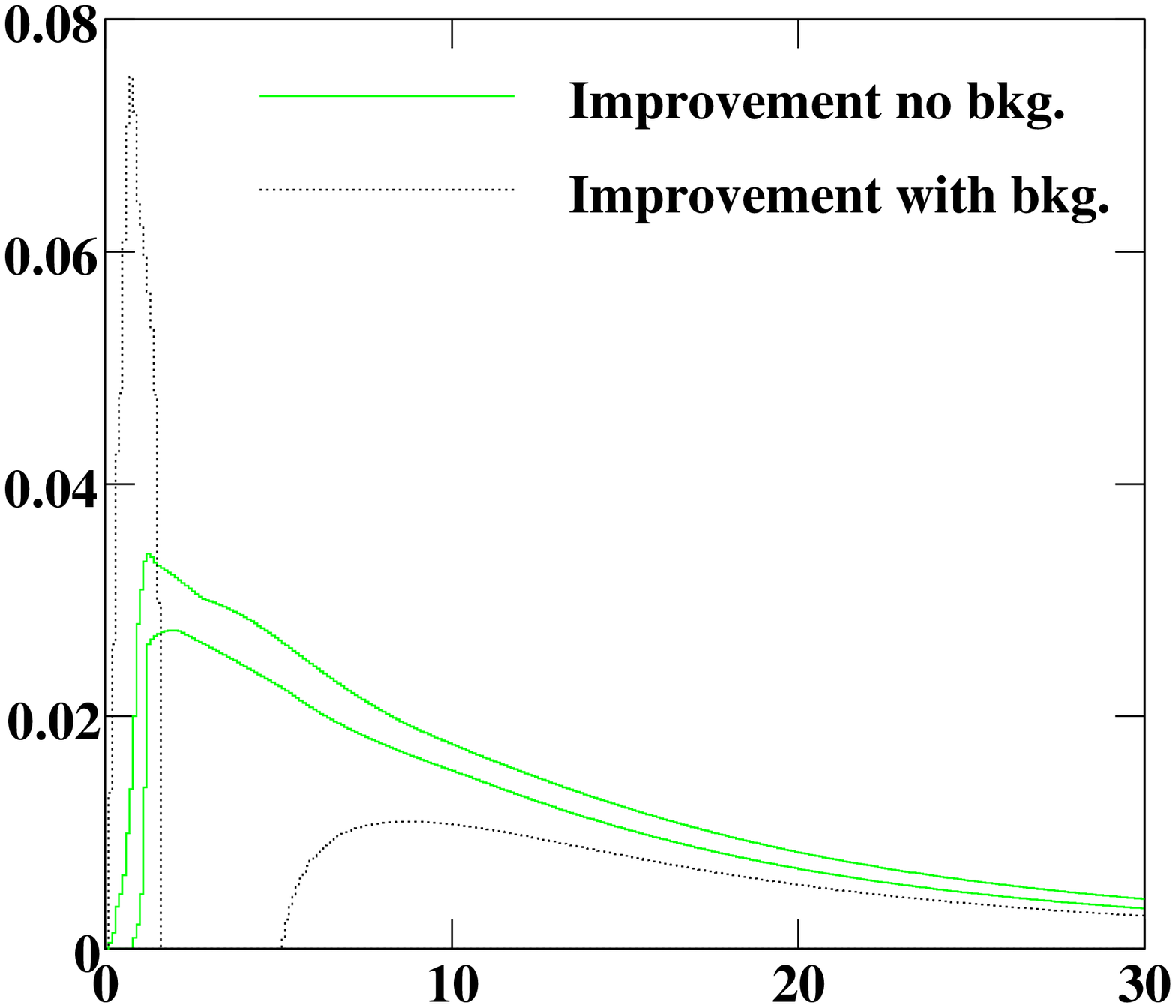}
\caption{Normalization relative improvement limits as a function of $\lambda_N$, without background or with a 
background twice less (left), twice larger (middle) and four times larger (right) than usual. 
In the right figure, only the upper limit of the normalization improvement with background is visible.
These improvement values were obtained after, at most, 50 steps.}
\label{fig:optLnOB}
\end{center}
\end{figure}
The optimization of the parameter $\lambda_N$ used in the normalization procedure, was studied for the 
comparison between data and reconstructed MC, with a null estimate of the remaining background fluctuations.
As explained before, at this level, one would not take advantage on the corrections of the resolution 
effects on new resonances.
Actually, these corrections make easier the separation of new structures and remaining background 
fluctuations.
A procedure that works properly at the data/reconstructed MC level, would work even better when comparing
an intermediate unfolding result and the true MC.
Furthermore, the normalizations at these two levels are improved by a better knowledge of the remaining 
background fluctuations and by the improvement of the transfer matrix through the true MC.

The first part of this study was performed with toys of data and reconstructed MC spectra, obtained as
described in Section \ref{subsec:building1}, with or without the remaining background given by 
Eq. (\ref{eq:bkgd}).
In both cases, the first estimate of $NE_{dSmc}$ is given by the number events in the data $NE_d$.
For 100 toys, we plot the minimal and maximal improvement brought by the procedure ($(NE_d- NE_{dSmc})/NE_d$), 
for a large interval of $\lambda_N$ values (see Fig. \ref{fig:optLnUB}).
We stop iterating when the relative improvement of two consecutive steps gets less than $10^{-6}$, or after
a fixed number of steps (if the first condition was not fulfilled before).

Fast convergence and good stability (with respect to the background problem) are exhibited for large values of
the parameter of normalization.
The procedure is clearly sensitive to the presence of background fluctuations, which are treated more like
real structures, for small values of $\lambda_N$.
It is only in this region of values of $\lambda_N$ that the limit on the number of steps plays a role.
Too small a value of $\lambda_N$ could be identified, even when running on fluctuating data, by the 
presence of instabilities in the sign and size of the correction (their size could not converge for a reasonable
number of iterations).
One can conclude from these plots that a value of $\lambda_N \approx 5$ would provide a relatively good 
normalization improvement at the first steps of the unfolding, as well as when one gets an estimate of the
background fluctuations.
Other amplitudes of background fluctuations (see Fig. \ref{fig:optLnOB}) could yield to different conclusions.
If these fluctuations are larger or not well known, one could prefer to use a large value of the parameter 
(which could even come to perform just the standard normalization) before the first background estimation, 
and a smaller one afterwards.
It has been tested that a change of the centre of the background distribution (Eq. (\ref{eq:bkgd})) by $\pm \sigma_b$ 
does not change the conclusions related to the normalization parameter.

In practice, one can perform the previous study on a model similar to the data or directly using the data and
reconstructed MC distribution.
In this second case, only the black curve in Fig. \ref{fig:optLnUB} would be available.

The plots in the rest of this subsection were obtained for:
\begin{itemize}
\item
an amplitude of the resonance initially centred on the bin $90$ (moved to the bin $87$, with a width 
$\sigma_1 = \sqrt{5}$ for this scan), reduced by $25\%$ with respect to Eq. (\ref{eq:bias1});
\item
the amplitude of the resonance initially centred on the bin $170$ (centred on the bin $173$, with a width
$\sigma_2 = 5$ for this scan), increased by $25\%$;
\item
the effect of the background subtraction reduced by $25\%$ with respect to the one in Eq. (\ref{eq:bkgd}), 
and centred on the bin $37$, with $\sigma_b = 5$.
\end{itemize}
The resulting optimal parameters are stable with respect to various changes of this type.

The choice of the value $\lambda_L$ (to be used for the first unfolding step) is made such that, for different 
relative amplitudes of the subtraction effect, it is not unfolded, and hence does not bias
other regions of the spectrum.
We have observed that the value $\lambda_L \approx 7$, would be large enough, not to unfold fluctuating background 
effects for different relative values of their amplitude.
In order to show that this parameter can be dropped, and the procedure simplified, we have however choosen a 
very large value for it, such that the correction done by the first unfolding is given only by the difference 
between the true and reconstructed MC.

\begin{figure}[h]
\begin{center}
\includegraphics[width=5cm]{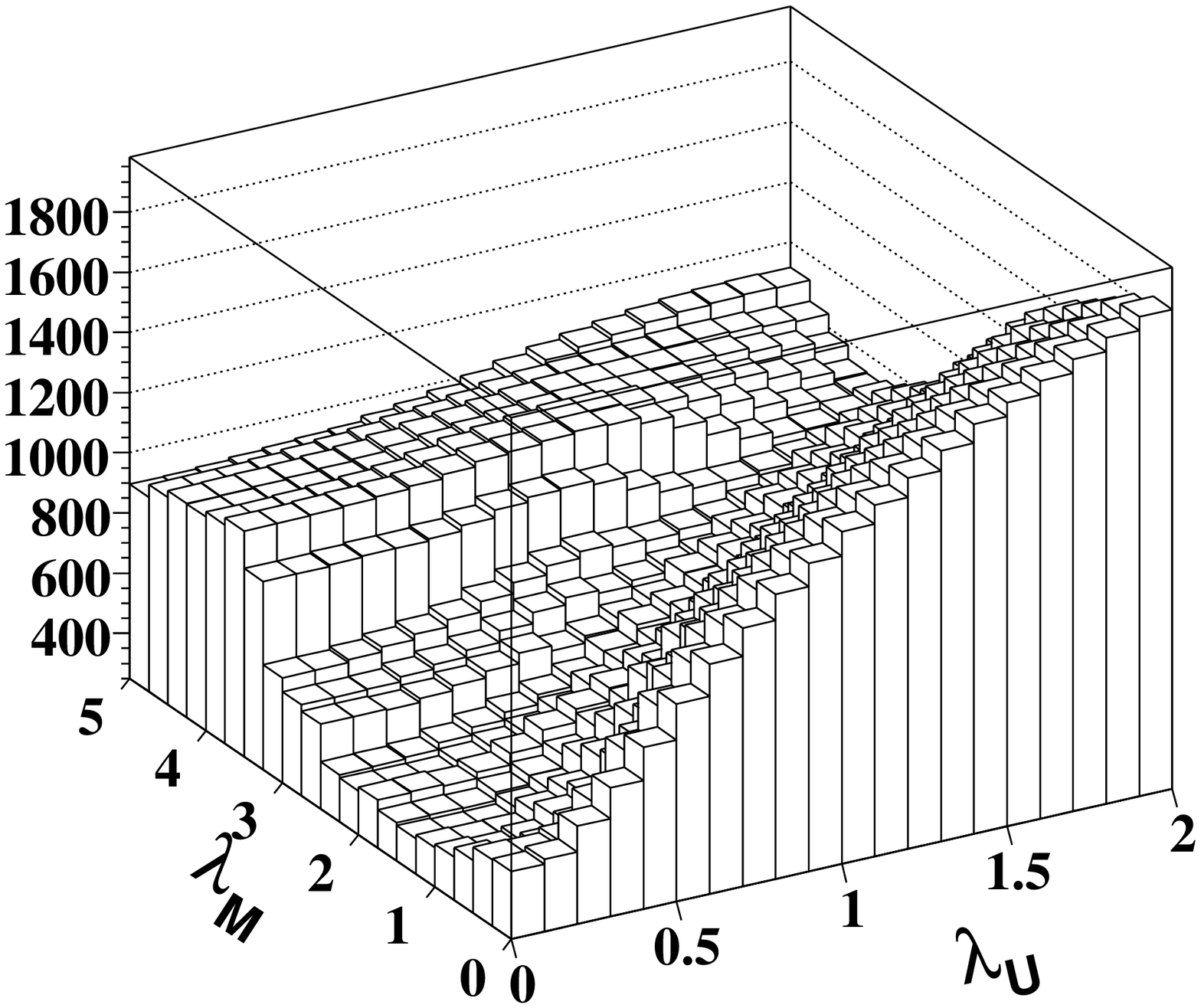}
\includegraphics[width=5cm]{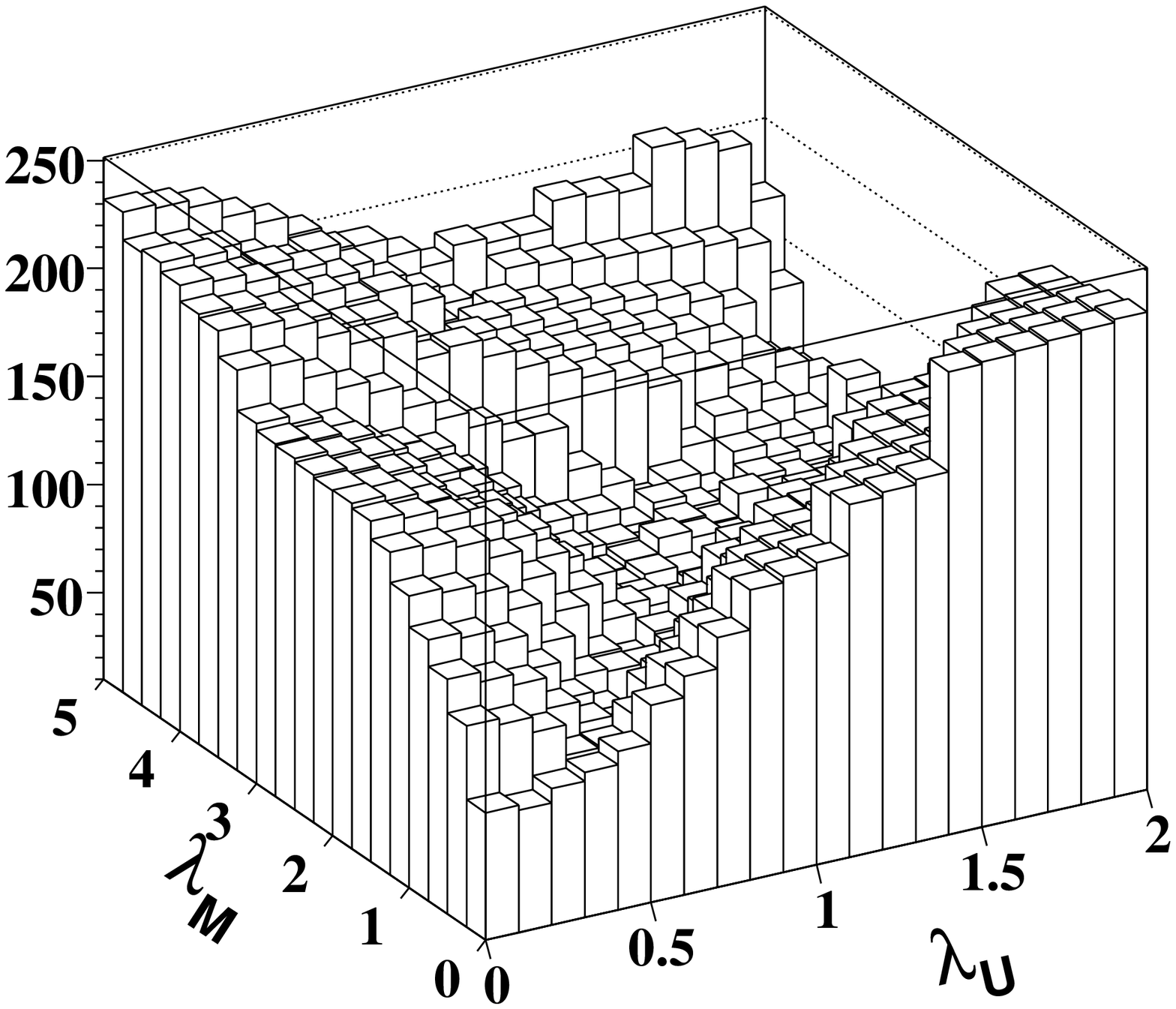}
\includegraphics[width=5cm]{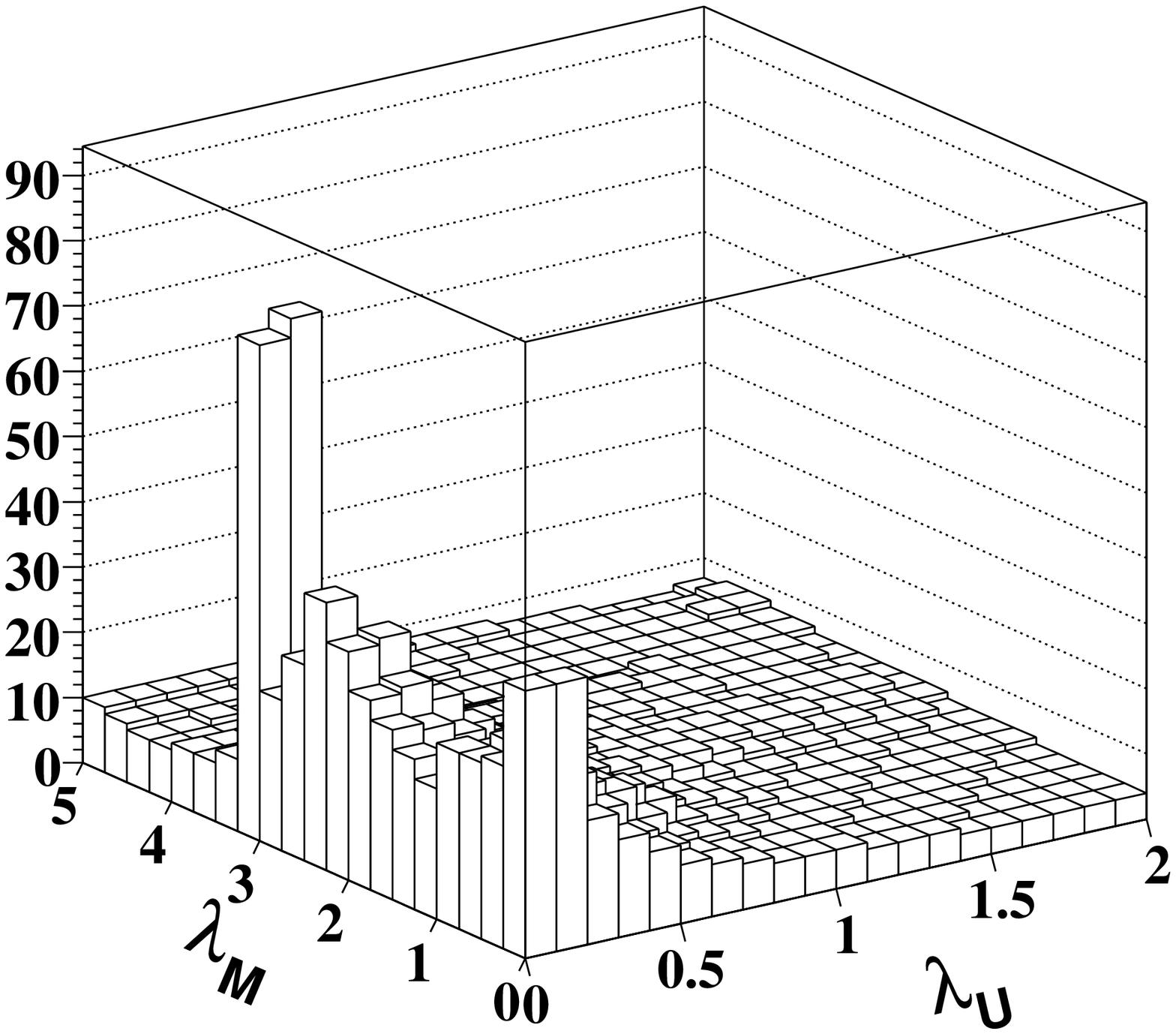} \\
\includegraphics[width=5cm]{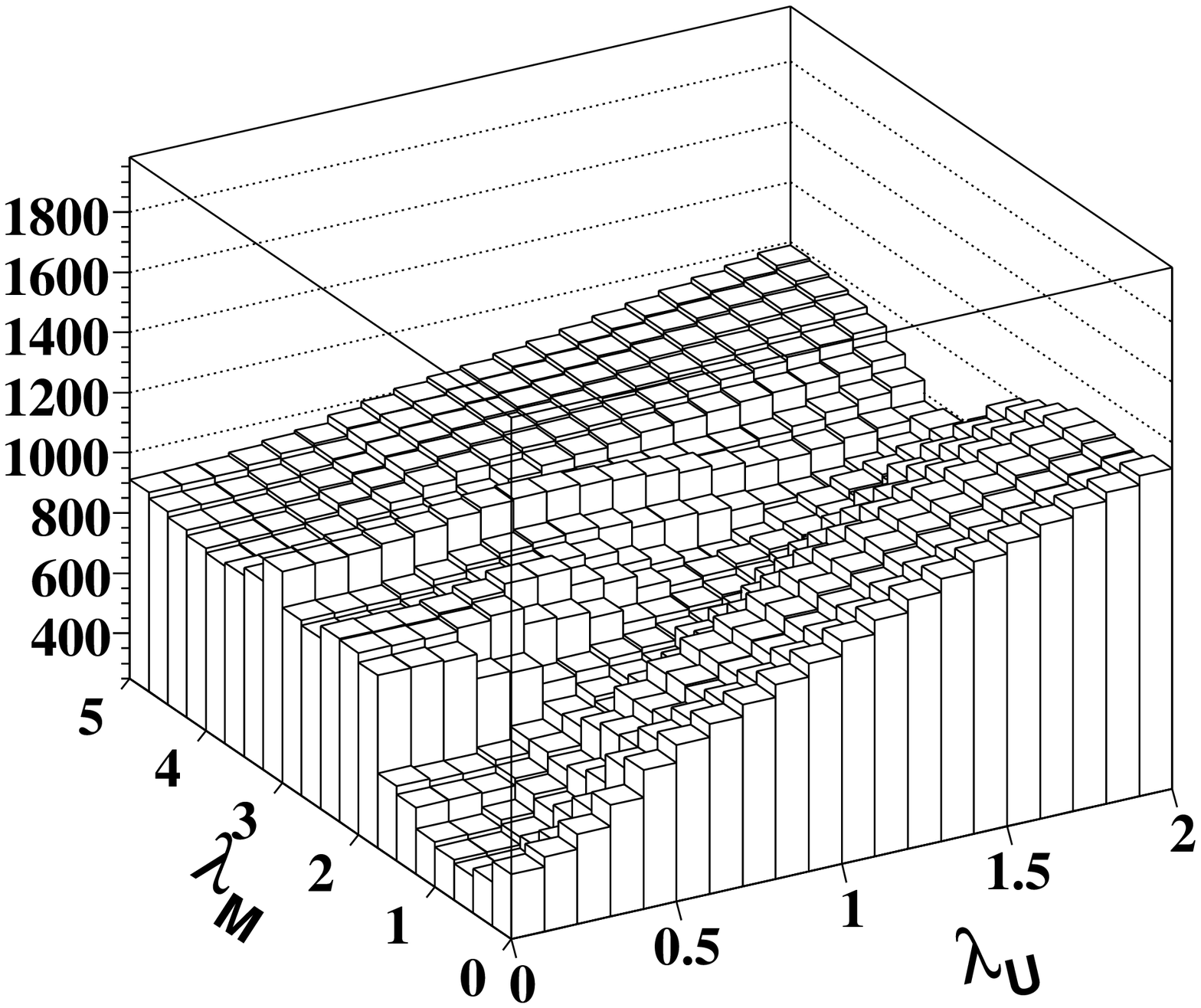}
\includegraphics[width=5cm]{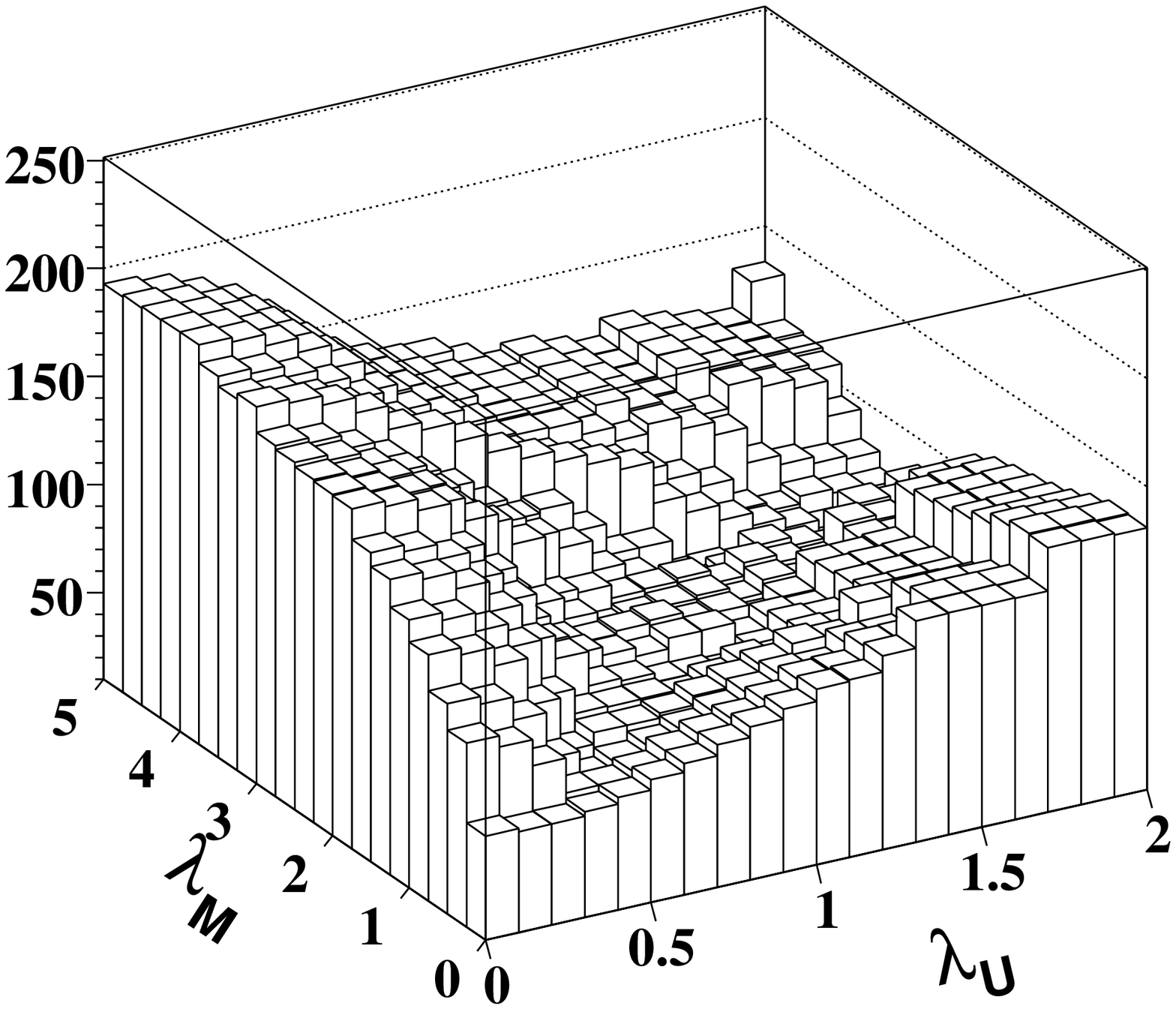}
\includegraphics[width=5cm]{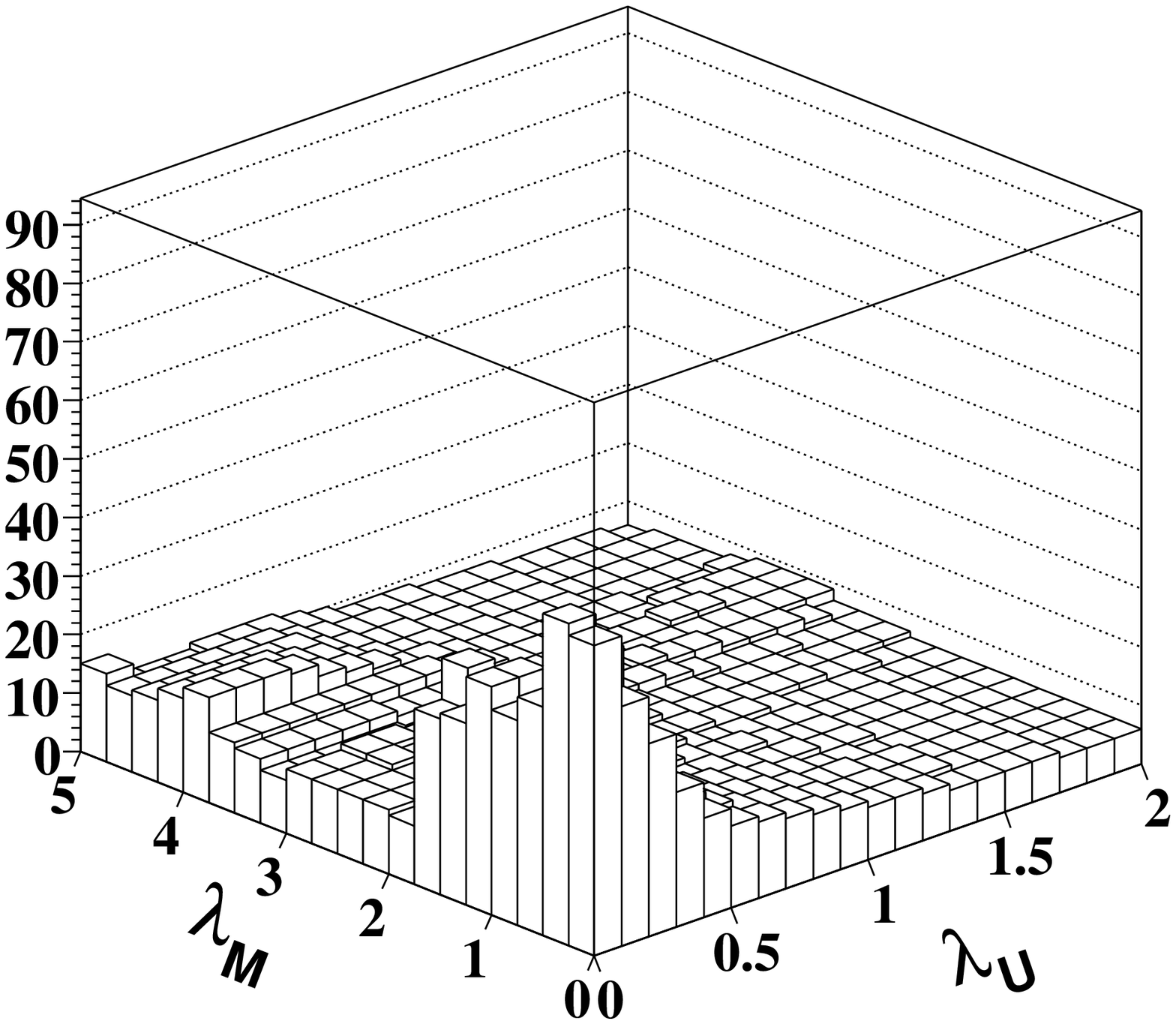} \\
\includegraphics[width=5cm]{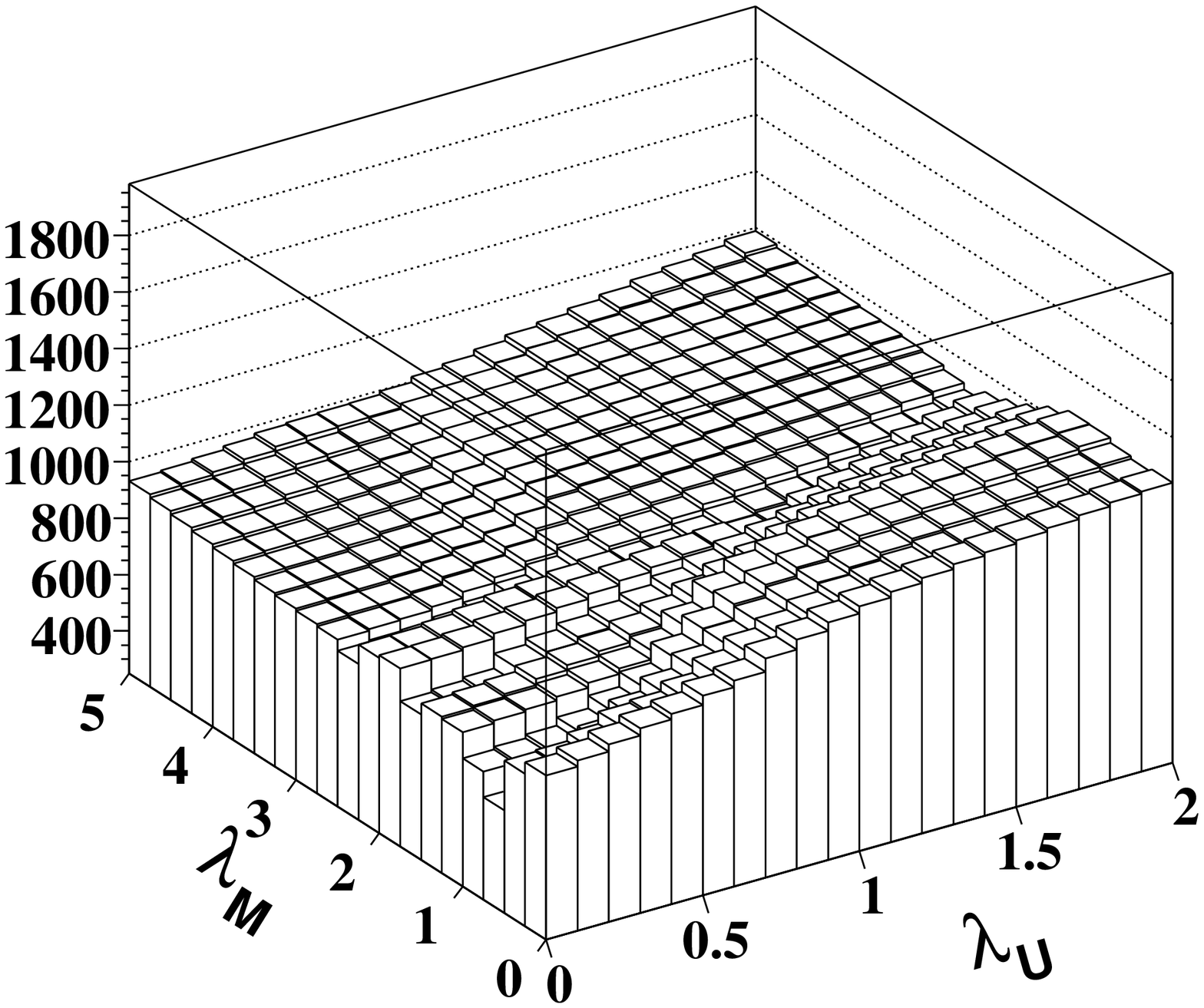}
\includegraphics[width=5cm]{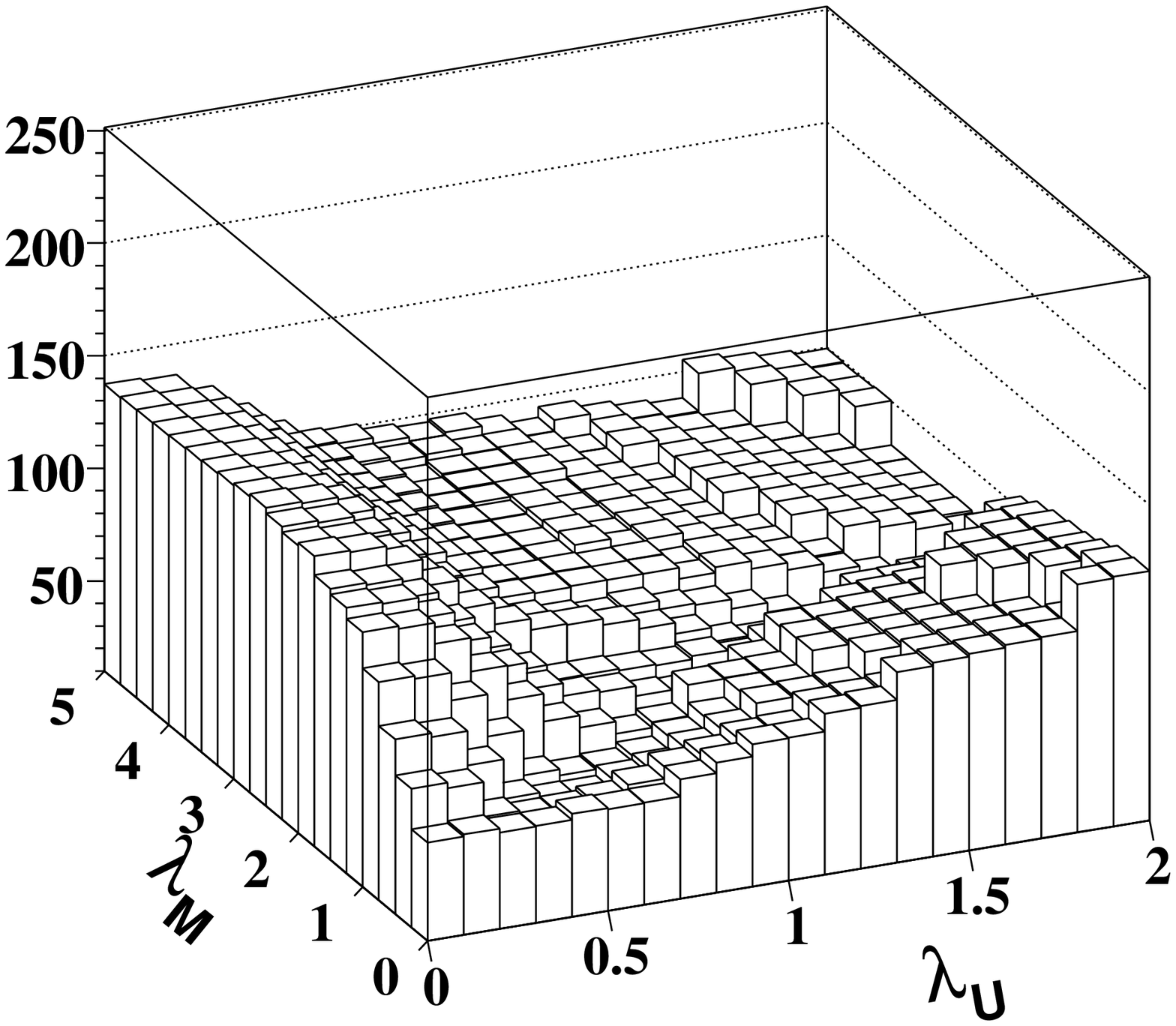}
\includegraphics[width=5cm]{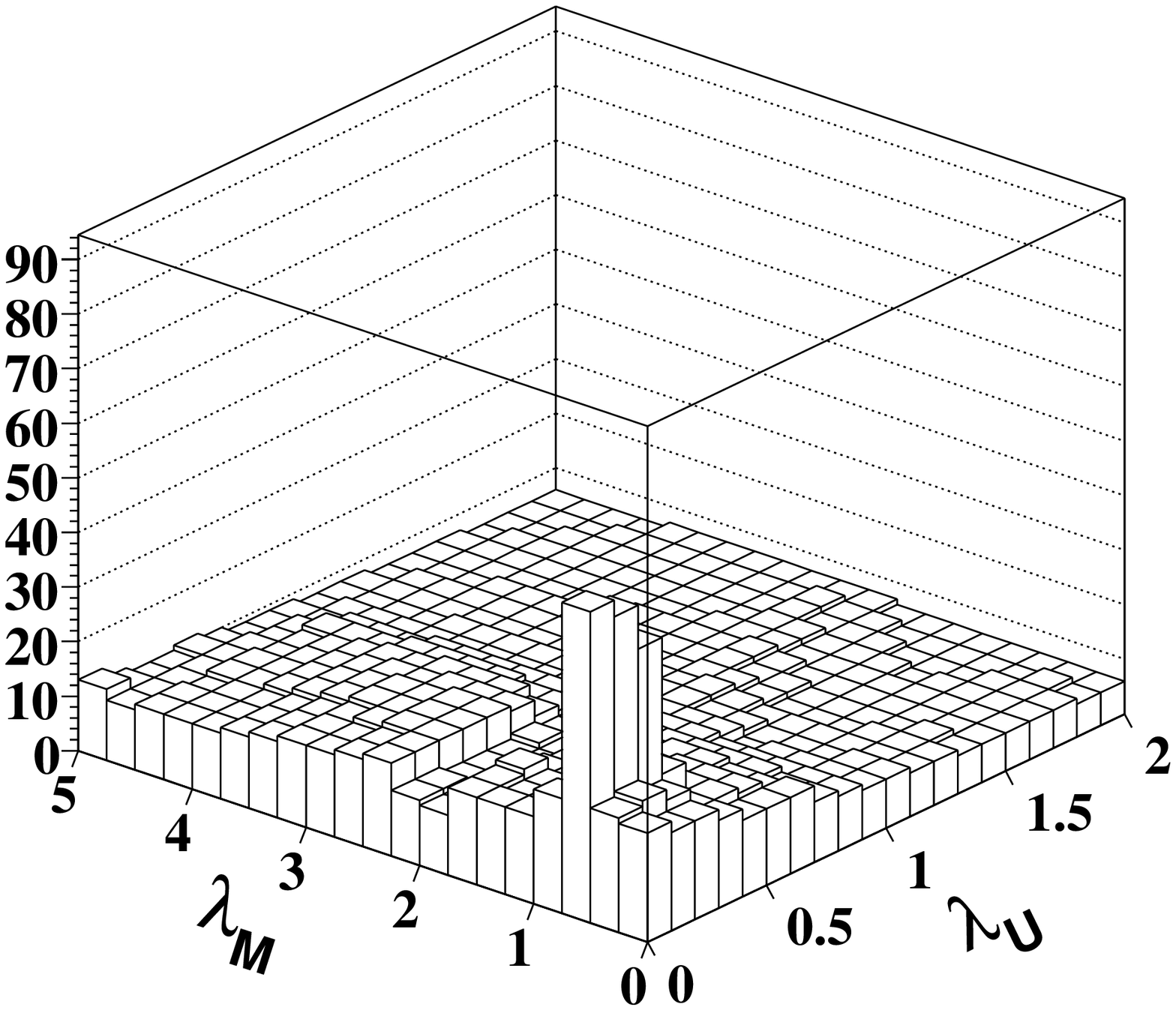}
\caption{Distance between the unfolding result and the true spectrum plus background fluctuations (left
column), distance between the final reconstructed MC and the data minus the events to be subtracted (middle
column), and the optimal number of steps (right column).
These plots were obtained for $\lambda_S = 3$ (top line), $\lambda_S = 5$ (middle line) and 
$\lambda_S = 7$ (bottom line).}
\label{fig:optLuLm}
\end{center}
\end{figure}

The two other parameters to be used and the number of necessary iterations, are analysed using 
a scan of their values, for test spectra with different amplitudes of the initially unknown resonances.
For every set $(\lambda_M, \lambda_U)$, we compute a distance of the $\chi^2$ type between the
final reconstructed MC and the data minus the events to be subtracted, as well as between 
the unfolding result and the true spectrum plus background fluctuations, for a number of steps 
minimizing this last distance (see Fig. \ref{fig:optLuLm}).
These distances are computed using simply the statistical errors of the data, plus a small component 
for statistical fluctuations of the MC, for the first one.
The anomalous values (once divided by the number of degrees of freedom), obtained sometimes for 
the optimal parameters, can be explained by the fact that we ignore the change in the errors 
of the spectrum and they correlations due to the unfolding.
However, as we are not interested in the absolute values of these distances, but only on 
the optimal parameters, this way to proceed is good enough.
Actually it is even to be preferred due to its inherent sensitivity to some potential instabilities of
the iterations (for inappropriate values of the parameters), generating large fluctuations.
This sensitivity would be lost if one would compute the distances using for example the errors resulting
from a toy simulation.

Asking for small values of the two distances (at the unfolded/true MC and respectively data/reconstructed MC level), 
one can clearly discriminate the best region of values for $\lambda_M$ and for $\lambda_U$.
They are valid, not only for the parameters of the spectra used in this scan, but also for different 
ones used in the test of the method, as we'll see in the following.
The corresponding plots indicate here small distance values for $\lambda_U$ and $\lambda_M$ close to zero 
(or even null), indicating that the use of these parameters is not absolutely necessary.
These plots also give even more information about potentially problematic parameterizations yielding for example 
badly unfolded (not enough corrected) new structures if $\lambda_M$ is too large, or to fluctuations if one 
performs only small unfolding corrections in the iterations (large $\lambda_U$) and the other parameters preventing their 
propagation are not appropriate.
This kind of study has some sensitivity to $\lambda_S$ (to be used in the subtraction procedure) too, through the distance 
at the unfolded/true MC level. 
Actually, if its value is too large, the subtraction is too strong preventing the good reconstruction
of new structures.
As we saw before, this distance is also enhanced if fluctuations propagate (for example because of a too small $\lambda_S$).
These distances provide however only global comparisons of the spectra. 
Therefore, for complex situations like in this example, a more precise determination of the parameters is provided by
local comparisons of the spectra within toy simulations (i.e. local contributions to the previous distances).
This kind of comparison performed for the spectra used in the previous scans, or other similar variations, indicates
that a value $\lambda_S \approx 5$ allows one to fix background fluctuations and to prevent the ones of the procedure,
without a bias of the reconstruction of new structures.
As we will see, the choice of this value, as well as the other parameters, can be tested directly when unfolding the data.

The plot of the optimal number of iterations provides a relatively good estimation of the number 
of steps which are needed in practice, when the corresponding unfolding parameters are used.
The confidence one can make to this number depends however on the precision with which one knows the
widths of the new resonances.
One should keep in mind that this is just an estimation and, once the unfolding parameters are determined,
a better suited strategy could be to stop the algorithm when the iterations start having a small impact on the result or
yet when one gets a good agreement between data and reconstructed MC.

\subsection{Unfolding the spectrum, possible tests of the result and systematic errors}

Here we present the way the unfolding behaves for the distributions described in Section 
\ref{subsec:building1}.
This tests the reliability of the unfolding method as well as the one of the procedure used to
optimize the parameters.
 
\begin{figure}[h]
\begin{center}
\includegraphics[width=16cm]{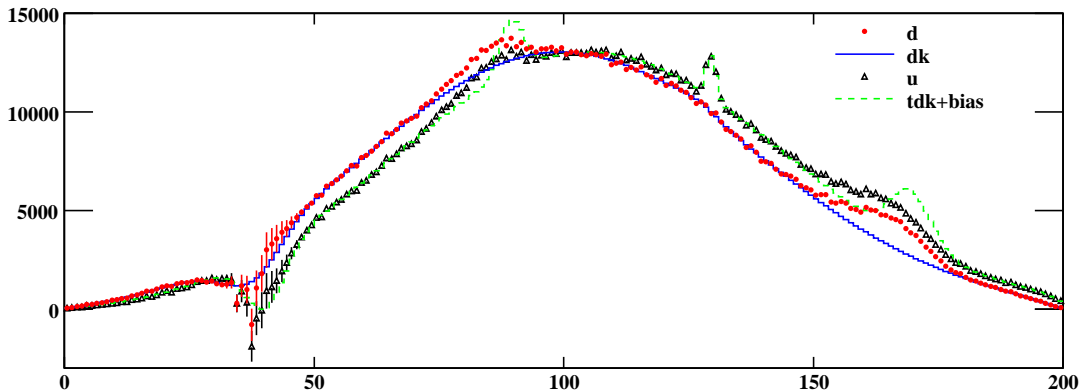}
\caption{The unfolding result after the first step (triangles), compared to the data distribution (filled circles),
the data distribution known by the MC (solid line) and the true data known plus the bias (dashed line).}
\label{fig:Unf1}
\end{center}
\end{figure}

\begin{figure}[h]
\begin{center}
\includegraphics[width=16cm]{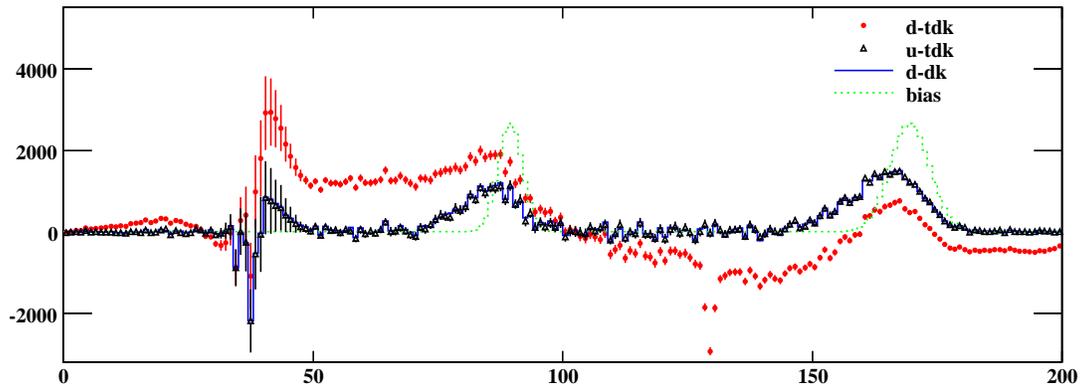}
\caption{Unfolding result after the first step minus true data known by the MC (triangles), data minus true data known 
by the MC (circles), data minus data known by the MC (solid line), and bias (dotted line).}
\label{fig:test1}
\end{center}
\end{figure}

As explained in the previous subsection for the various test distributions, here also a parameter 
$\lambda_L \approx 7$ would be large enough to leave the background contribution in its initial position.
The first unfolding step was however performed with a very large value for this parameter 
(see Fig. \ref{fig:Unf1} and \ref{fig:test1}) and it corrects all the elements of the spectrum which are 
known by the MC, for both kinds of transfer effects (in spite of the fact that they are relatively important).
This first step could also perform part of the needed corrections for the new structures, but we postpone
these corrections to further unfolding steps, by taking a very large $\lambda_L$.

A series of 65 iterations was then performed, with complete unfolding and matrix modification (i.e. taking
the parameters $\lambda_U \to 0$ and $\lambda_M \to 0$).
Actually, in this case the regularization is provided by the use of $\lambda_S = 5$, and the use of
$\lambda_U$ and $\lambda_M$ can be dropped.
A very good estimation of the remaining background fluctuations is obtained at the first iteration, and 
the further improvements are relatively small. 
However, for the sake of generality, it has been tested that good results can be obtained, with non-zero values 
of $\lambda_U$ and $\lambda_M$.

\begin{figure}[h]
\begin{center}
\includegraphics[width=10cm]{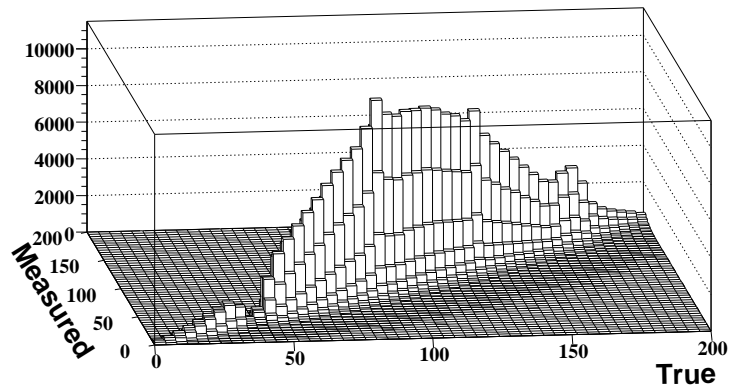}
\caption{Transfer matrix after 65 iterations.}
\label{fig:Amx4Nsteps}
\end{center}
\end{figure}

The main effect of these iterations is to correct the folding matrix, by introducing in the true MC 
distribution (and implicitly in the reconstructed one), the structures which were unknown for 
the initial simulation (see Fig. \ref{fig:Amx4Nsteps}). 
A first reason for which such a high number of iterations was needed in this test are the large resolution
effects which must be corrected for the initially unknown structures, even for the narrow one.
It has been tested that between 15 and 20 iterations are generally enough to reconstruct well the new structure 
centred on the bin 170, which is slightly larger than the detector resolution, being however strongly affected 
by its effects too.

\begin{figure}[h]
\begin{center}
\includegraphics[width=16cm]{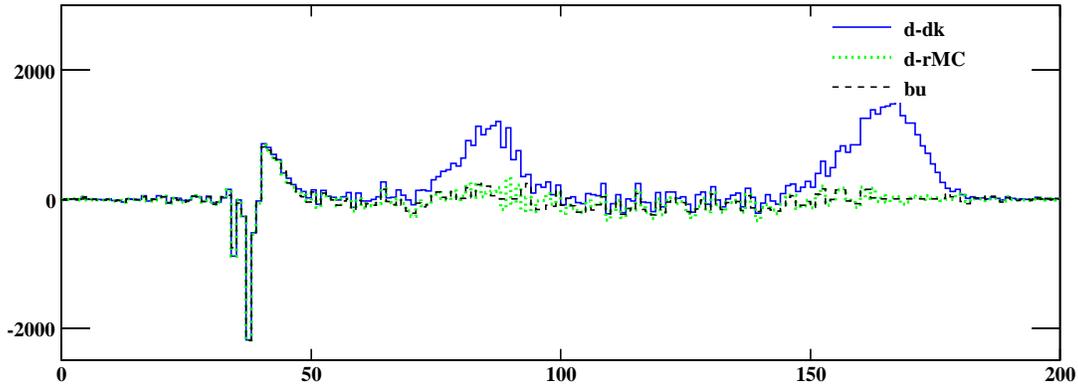}
\caption{Remaining background estimate (dashed line), data minus the improved reconstructed
MC (dotted line) after 65 iterations, and unknown data for the initial MC (solid line).}
\label{fig:data1-recoNsteps}
\end{center}
\end{figure}

The final estimation (see Fig. \ref{fig:data1-recoNsteps}) reproduces well the background fluctuations
added in the region of the dip, for the bins where it was fluctuating as well as for the ones where it
was stable.
The need to distinguish between potentially remaining background fluctuations and new structures, especially 
when having an important transfer of events in the unfolding, imposes important constraints on the parameter 
values that can be employed.
This can be a the second reason which can increase the number of needed iterations.
The difference between the data and the improved reconstructed MC is basically equal to the background 
estimate, which is not used when modifying the transfer matrix.

\begin{figure}[h]
\begin{center}
\includegraphics[width=16cm]{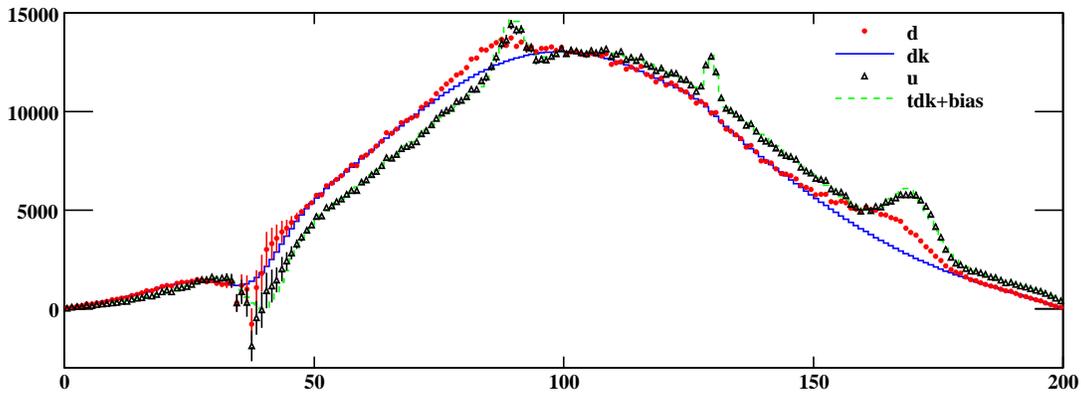}
\caption{The unfolding result after 65 iterations (triangles), compared to the data distribution (filled circles),
the data distribution known by the MC (solid line) and the true data known plus the bias (dashed line).}
\label{fig:toyUnfoldingN1}
\end{center}
\end{figure}

\begin{figure}[h]
\begin{center}
\includegraphics[width=16cm]{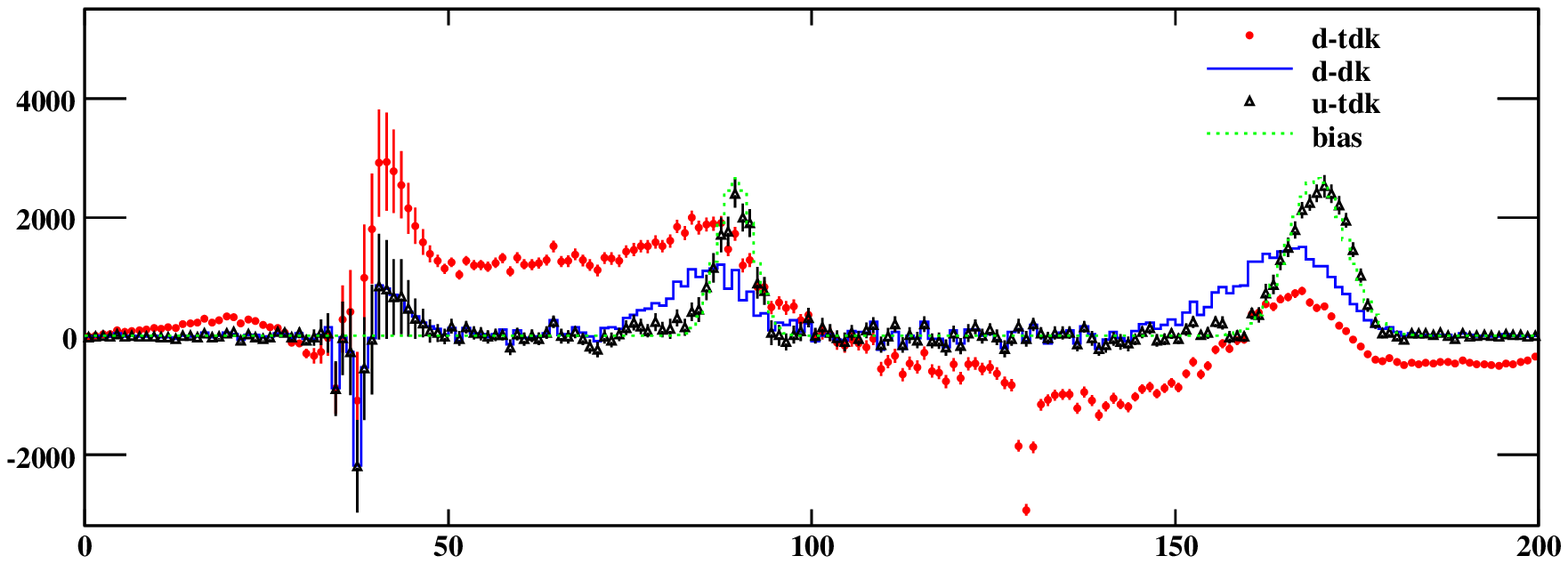}
\caption{Unfolding result after 65 iterations minus true data known by the MC (triangles), data minus true data known 
by the MC (circles), data minus data known by the MC (solid line), and bias (dotted line).}
\label{fig:testNsteps}
\end{center}
\end{figure}

The final unfolding result reconstructs well all the structures in the true data, without introducing
important systematic effects due to the remaining background (see Fig. \ref{fig:toyUnfoldingN1} and Fig. 
\ref{fig:testNsteps}).
The error bars of the unfolding result(s) were estimated using 100 MC toys, with fluctuated data and 
transfer matrix for the unfolding procedure.
The corrected structures are affected by fluctuations coming from all the data bins contributing to them,
which is particularly important for large folding/unfolding effects like the ones in this example.
As expected, new structures are more affected than the ones initially known by the MC, because the corresponding
data fluctuations can locally affect the improved MC transfer matrix.

At this point one could perform a test of the parameterization of the procedure, directly on the data.
Actually one could perform a series of toy simulations of the unfolding, and compare the average
of the unfolding results with the result found previously.
A significant difference between the two, in any region of the spectrum, would indicate a problem in
the parameterization, which should be reconsidered.
An increase of the fluctuation of the unfolding result in the toys with respect to the ones of the data, 
comparable or larger than the correction performed on new structures (i.e. by iterations if the first step 
is trivialized as before), would indicate the same kind of troubles.
No such problems were observed in the previous example.

If the study described in the previous subsection, using global distances between the spectra, is not enough 
sensitive to the values of the unfolding parameters, one can always build toy models for the data (following 
\ref{subsec:building1}) and perform the unfolding as explained before, in order to tune the method.
One can use in the folding the matrix given by the MC, while a statistically fluctuated one will be used
in this unfolding.
The bias, used to get the true data, and the remaining background fluctuations, will be choosen such that 
the folded toy data are qualitatively close to the real ones.
The local comparison of the unfolding result with the toy true data provides an excellent sensitivity to the
parameters of the method.
Once the method is tuned, the possible (small) remaining difference between the unfolding result and the true
data can be used to estimate the systematic error of the method.
This difference could even be used to compute a correction of the result, provided that it is stable with respect
to statistical fluctuations, the uncertainty of the data true model is smaller than this difference, and if one 
can not further improve the parameters of the method.

Among the functions listed in Eq. (\ref{f:regFunctions}), (\ref{f:1Mexp}) and (\ref{f:1Mfrac}), the ones with a null derivative at the origin 
seem to provide a better separation of fluctuations and real deviations.
The small differences between the results obtained with the several (well suited) functions, as well as
the ones related to some reasonable changes in the regularization parameter(s), can be used to estimate 
the corresponding systematic errors of the procedure.
There is however at least a partial overlap between this estimation of the systematic error and the one proposed
in the previous paragraph.

The systematic errors related to the unfolding are not only dues to the procedure, but also to the knowledge of
the folding matrix.
The difference between data and MC at this level can be estimated in practice by detailed studies of events
affected by effects that the unfolding procedure must correct.
If the amount of some of these events is different in data and in the simulation, one can even correct the last one.
This could also reduce the correction the unfolding has to bring at the level of the spectra.
One can estimate the uncertainty on the transfer matrix coming from this (or yet another) kind of effects/corrections.
Performing the unfolding with the transfer matrix modified within these uncertainties allows one to compute
the systematic error associated to the final spectrum.
There can indeed also be an overlap between this estimation of the systematic error due to the transfer matrix
and the ones previously associated to the unfolding procedure.

We have seen that this unfolding method allows one to unfold structures having a width smaller than the detector
resolution.
However, these reconstructed structures are for sure affected by the fluctuations of the data bins where their events
were initially situated, and by the precision of the knowledge of the transfer matrix.
These two effects put a limit on the quality of the reconstructed structures.

The use of the regularization functions in this method imply non-linear effects in the passage from data to the
unfolded spectrum.
This is actually a general feature of the regularization methods used in the unfolding procedures.
It can slightly affect the shape of the error distribution of the unfolded spectrum.
In this example, this type of effect was observed mainly at the edges of the new structures introduced in data, where
it is more difficult to distinguish between statistical fluctuations and real significant deviations.

\section{A simpler example for the use of the unfolding procedure}
\label{Sec:SimpleEx}
In this section we provide an example that could be closer to the usual applications of this method, namely
when the effects of the folding and the data-MC differences are relatively small.
We indicate how the method can be considerably simplified for this kind of application and how one can easily
build a test (or a procedure to set the remaining parameter(s) ) of the method.

\subsection{Building the data spectra and transfer matrix}
\label{subsec:building2}
To build this second example we use the same procedure as the one described in subsection \ref{subsec:building1},
but we bring important changes in the spectra and transfer matrix.
In order to provide a rather low statistics test of the method, we reduce the global statistics of MC and data by a
factor of about 20 (their relative ratio is equal to 10, like in the previous example).

Concerning the transfer matrix, besides this change, we remove the dip around the bin $40$ in Eq. (\ref{eq:Tmatrix})
and we reduce resolution effects by taking $\sigma_y = 1$.
The effect of systematic transfers of events from high index bins to the lower ones is also reduced by putting $c_1 = 8$ 
and $c_2 = 2$, in Eq. (\ref{eq:Amatrix}).

The ``bias'' of the true data with respect to the true MC is changed to a smaller and a smoother one:
\begin{equation}
bias_j = \frac{75}{\sqrt{2\pi}} \cdot
	\left[ - e^{-\frac{(j-50)^2}{2\cdot 20^2} } 
	+ e^{-\frac{(j-140)^2}{2\cdot 20^2} } \right] ,
\label{eq:bias2}
\end{equation}
and we do not introduce any remaining background fluctuations in this example.
The errors associated to the data points are simply the statistical ones.

\subsection{A simplified version of the unfolding procedure}
Given the fact that the complexity of this test is considerably reduced, with respect to the previous one, a
simplified (next to trivial) version of the unfolding procedure allows one to perform the needed corrections.

First of all, as in this example there are no significant new structures in the data, the normalization 
procedure looses its sensitivity and does not indicate any possible improvement at this level.
Indeed, one can see this by making a scan of the possible $\lambda_N$ values with fluctuating data, as it was shown 
in the previous example.
In the present case, the improvement limits interval contains the null value, for any value of the $\lambda_N$ parameter.
One can therefore drop this procedure here, for example by setting $\lambda_N = 0$ (i.e. by using the standard 
normalization).

The possibility of subtracting remaining background fluctuation is also dropped here.
In practice one can take this decision by comparing the size of the data point errors with respect to the available
statistics, or equivalently using the errors introduced by the background subtraction itself.
The possible effect of this simplification in the unfolding strategy can also be tested directly on data, as explained in
the previous example.

We will finally use a simplified unfolding procedure using a first step with the initial transfer matrix and a parameter
$\lambda_L$, and an iteration where the transfer matrix is improved and a new unfolding is performed.
In this example, one iteration modifying the matrix is enough to correct the systematic data-reconstructed MC differences.
\begin{figure}
\begin{center}
\includegraphics[width=16cm]{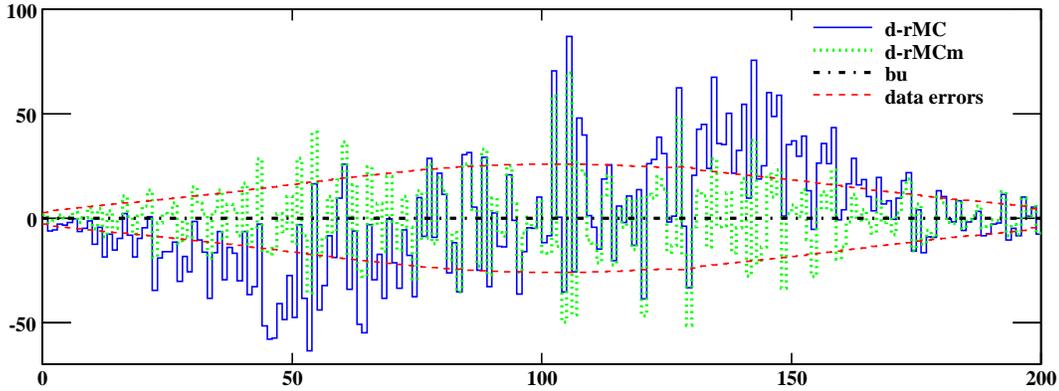}
\caption{Data minus data known by the MC (solid line), data minus the modified reconstructed MC (dotted line),
estimate of the remaining background fluctuations (dashed-dotted line), null in this example 
and the data errors (dashed line).}
\label{fig:SimpExUnf_data-recoMC}
\end{center}
\end{figure}
The figure \ref{fig:SimpExUnf_data-recoMC} indicates this difference before the unfolding, and after one iteration.
For the example shown here we chosed $\lambda_L = 1.5$.
Although this choice is not crucial for the effect shown in this figure, its motivation will become clear after the
following studies.
In order to measure the full potential systematic effect of this iteration, we set the corresponding parameters:
$\lambda_U = 0$ and $\lambda_M = 0$ (i.e. we perform the maximal change of the transfer matrix and the maximal unfolding 
correction afterwards).

\subsection{Building a simple toy test of the unfolding procedure and its parameterization}
We propose a direct test of the unfolding procedure and its parameterization, searching for potential systematic biases 
that could be introduced at this level.
For this test we build a toy true and reconstructed data, inspired directly by the original data spectrum and the MC
simulation (which could be directly used in practice).
The toy reconstructed data are then unfolded with a transfer matrix ($A'$) obtained after a fluctuation of the initial 
one available in practice ($A$).
Therefore, the MC for this test is identical to the one in the ``real'' unfolding, up to statistical fluctuations.
We compare the result to the toy true data.

We define the true data known ($tdk$) by the MC as the normalized initial true MC (obtained from the matrix $A$, without 
any additional statistical fluctuations).
We introduce the data ``known'' by MC, which are the result of the folding of the $tdk$, using the transfer matrix $A$.
They are identical to the reconstructed MC in the first unfolding step, up to statistical fluctuations between $A$ and $A'$.
As usual, we add a bias to the true data known by the MC, in order to get the ``true data'' ($td$).
In order to build a test as close as possible to the real situation, the bias is obtained from the difference between data
and the normalized initial reconstructed MC, multiplied by a constant factor.
The true data are then folded using the transfer matrix $A$.
We define the reconstructed data as the result of this folding, with or without final statistical fluctuations.
It is interesting to consider here these two variations of the test.
The first one, where the data are fluctuated statistically, is closer to the real unfolding operation and 
tests the existence of potentially spurious effects due to the limited statistics in the data (and MC).
The second test, allows an easier search for potential systematic effects of the method, which one can look for in 
the difference between the unfolding result and the true data.

\begin{figure}[h]
\begin{center}
\includegraphics[width=16cm]{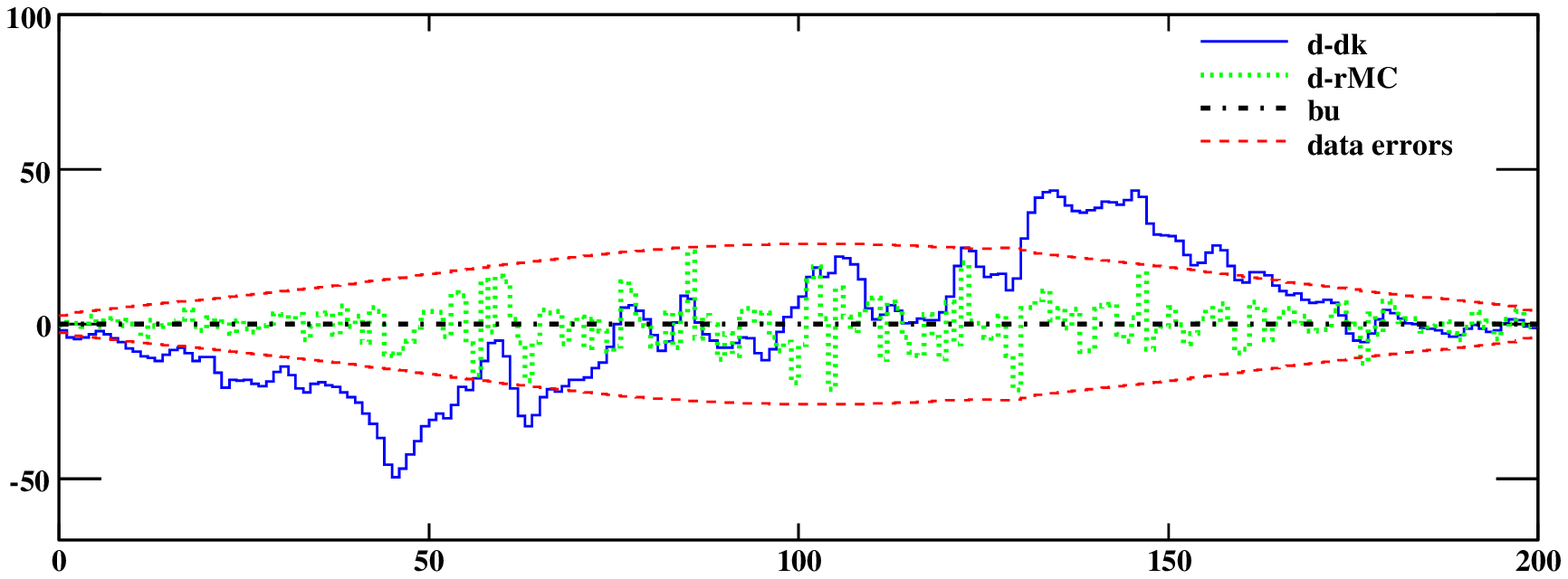}
\includegraphics[width=16cm]{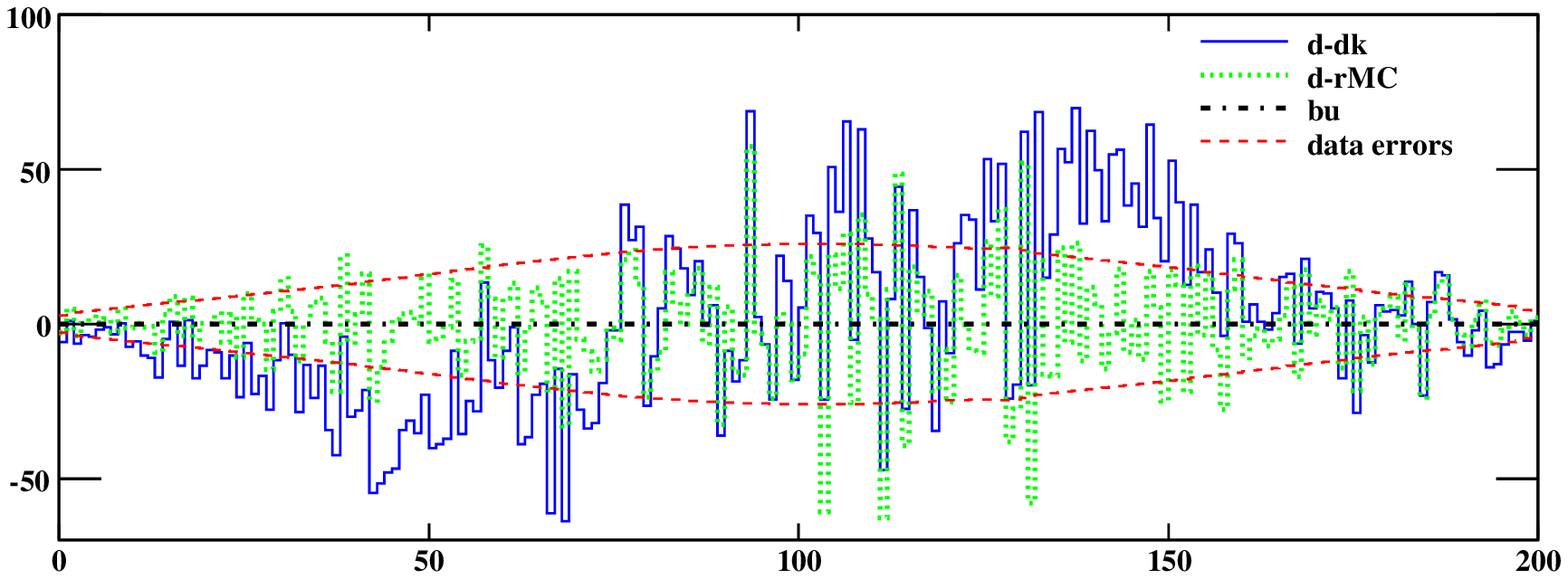}
\caption{Data minus data known by the MC (solid line), data minus the modified reconstructed MC (dotted line),
estimate of the remaining background fluctuations (dashed-dotted line), null in this example 
and the data errors (dashed line).
The upper plot corresponds to the case where no additional statistical fluctuations were added to the test data, while 
the data were fluctuated statistically for the example shown in the lower plot.}
\label{fig:SimpExTP_data-recoMC}
\end{center}
\end{figure}

If the multiplicative factor, used to obtain the bias, equals one, the data - ``known data'' difference in this test,
is very close to the data - reconstructed MC difference in the real unfolding procedure (up to final statistical fluctuations 
(if any), transfer matrix fluctuations and smoothing folding effects) (see Fig. \ref{fig:SimpExTP_data-recoMC}).
We consider this factor equal to one in the following test (if not stated otherwise).
Just as in the ``real unfolding'' (see Fig. \ref{fig:SimpExUnf_data-recoMC}) all the systematic effects in the 
data - reconstructed MC difference are corrected after the first iteration.
A test as the one build here is well suited, provided the data-MC difference and the unfolding corrections are relatively
small.

We measure the bias of the unfolding result with respect to the true data, after the first step and after one iteration.
\begin{figure}[h]
\begin{center}
\includegraphics[width=16cm]{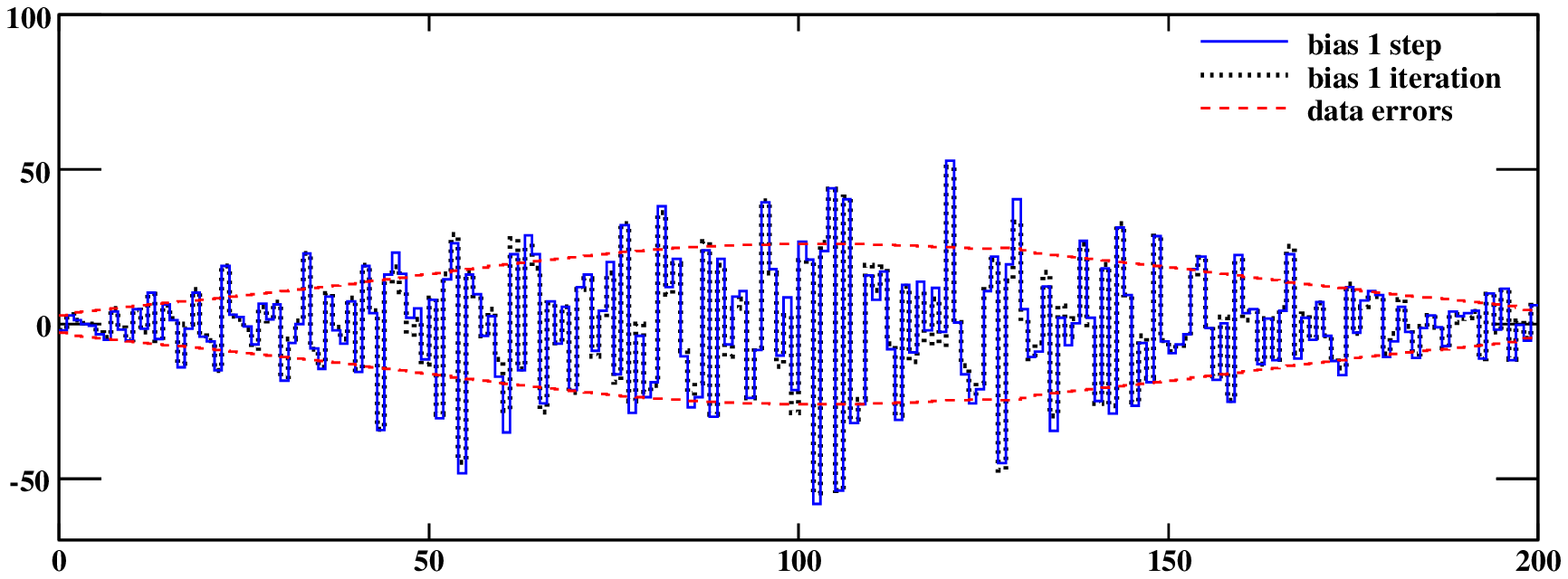}
\caption{Remaining bias after the first unfolding step (solid line), after one iteration (dotted line),
and the data errors (dashed line).
This plot corresponds to the case where no additional statistical fluctuations were added to the test data, and the first
unfolding was performed with $\lambda_L = 1.5$.}
\label{fig:SimpExTP_testBias}
\end{center}
\end{figure}
At this level one can not observe any systematic bias after the first unfolding step, and the impact of the iteration is
very small.

This study was performed for several sets of independently generated data.
In practice, one could build several models starting from statistically fluctuated data samples.
In order to measure the potential bias, we compute it in larger bins (each of which contains 40 initial bins).
For $\lambda_L$ values up to at least $1.5$ we find maximal relative biases in the interval $[0.1; 0.5]\%$.
This interval becomes $[0.2; 0.8]\%$ for $\lambda_L = 2$, $[0.3; 0.9]\%$ for $\lambda_L = 3$ and
$[0.8; 1.6]\%$ for very large $\lambda_L$ values (i.e. when no unfolding correction is done on the data-rMC difference).

This study shows that from the point of view of the potential bias remaining after the first unfolding step, the 
$\lambda_L$ values up to $1.5$ provide very comparable results.
However, the larger the $\lambda_L$ value, the less smoothing is done on the data, as the fraction of their fluctuations
kept in their initial bin for the final spectrum is larger, and the bins of the final spectrum are less correlated.

This type of test could also be used to identify situations where the detector transfer matrix is (almost)singular.
Such a situation would be encountered for example in the rather pathological situation of a detector with very bad resolution, 
in which case the data would be compatible with the $rMC$ no matter what $tMC$ distribution was used.
Therefore, the unfolding result would also be compatible with any other $tMC$ that was used to build the transfer matrix.
It is at this point that the test described in this subsection comes into play. 
Using various true data distributions for this test, one would assign a very large systematic error to the unfolding result,
making it compatible with any spectrum preserving the number of events.
This would of course indicate that the solution of the unfolding is not well defined.

\subsection{Unfolding the data}

\begin{figure}[h]
\begin{center}
\includegraphics[width=16cm]{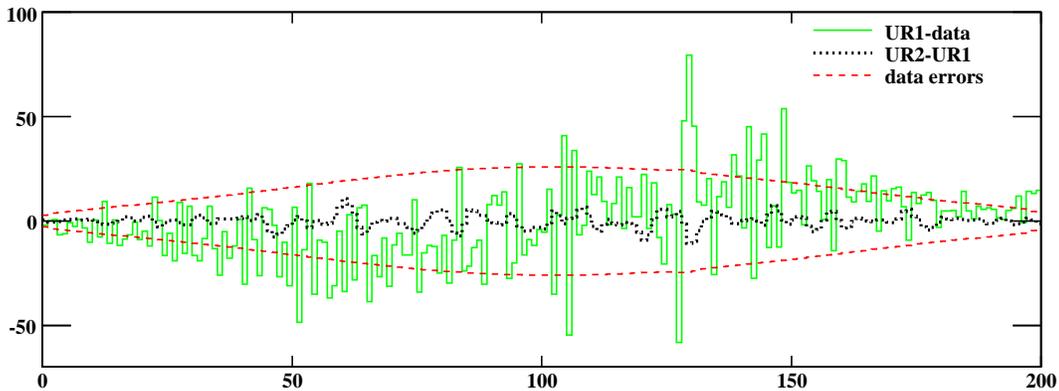}
\caption{Correction bringed by the first unfolding step (solid line), correction bringed by one iteration (dotted line),
and the data errors (dashed line).
This plot corresponds to the case where the first unfolding was performed with $\lambda_L = 1.5$.}
\label{fig:SimpExUnf_impIter}
\end{center}
\end{figure}

Just as in the case of the previous test, for the ``real'' unfolding, the first step provides the main systematic correction, 
while the effect of the iteration is very small (see Fig. \ref{fig:SimpExUnf_impIter}).
The impact of the first iteration on the unfolding result can be seen as another measurement of the systematic error
of the method.
Just as in the case of the toy test, we measured this effect in larger bins (each of which contains 40 initial bins).
We perform this study for several sets of independently generated data and transfer matrix.
For $\lambda_L$ values up to at least $1.5$ we find an averaged maximal (over 5 large bins) improvement of the order of
$0.1 \%$ with however important relative fluctuations between different data sets.
When increassing $\lambda_L$, the potential improvement gets larger and more significant with respect to fluctuations.
Its value is approximatively $0.2 \%$ for $\lambda_L = 2$, $0.5 \%$ for $\lambda_L = 3$, and $1.1 \%$ for very large
$\lambda_L$ values.
This confirms the previous conclusion concerning this test, namely that for $\lambda_L \lessapprox 1.5$ the first unfolding 
step provides already a good result, while the iteration(s) could be useful for larger $\lambda_L$ values.

\begin{figure}[h]
\begin{center}
\includegraphics[width=16cm]{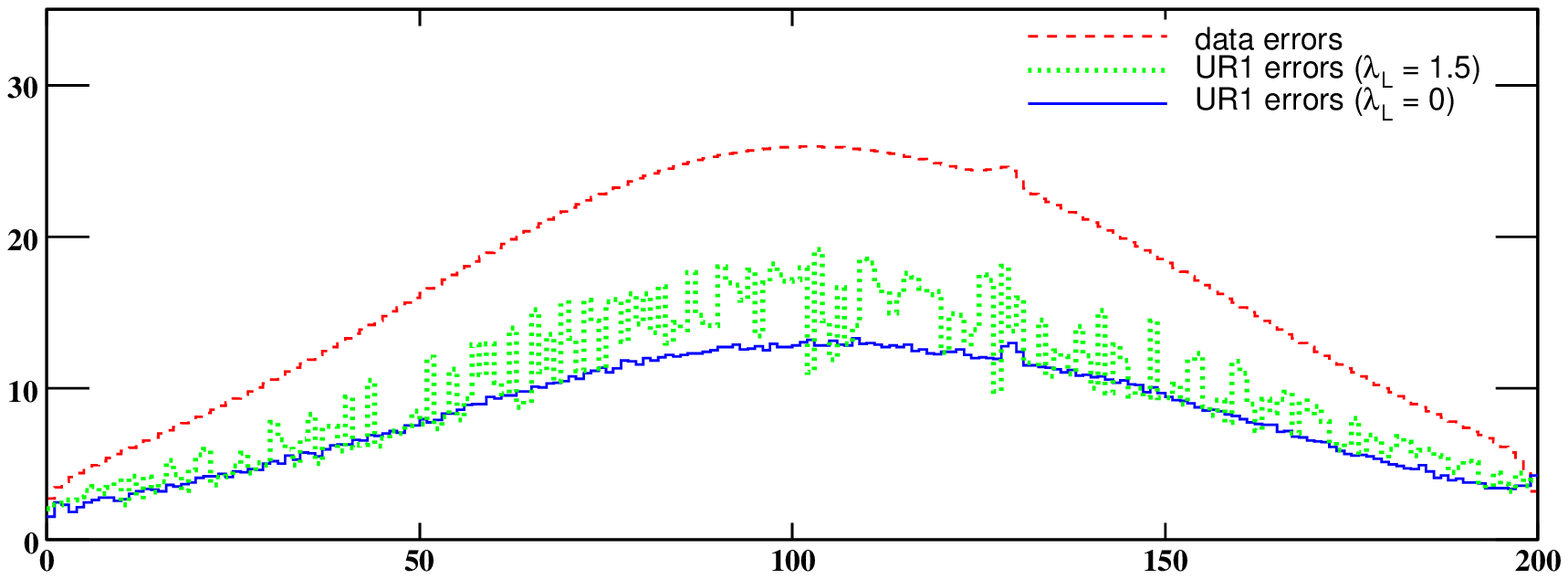}
\caption{Diagonal errors after the first unfolding step for $\lambda_L = 1.5$ (dotted line), for $\lambda_L = 0$ (solid line),
and the data errors (dashed line).}
\label{fig:SimpExUnf_errors}
\end{center}
\end{figure}

As explained before, using a larger $\lambda_L$ prevents the unfolding and smoothing of the data fluctuations, keeping the 
bins less correlated.
This effect can be observed at the level of the diagonal errors of the unfolding result, after one step (see Fig.
\ref{fig:SimpExUnf_errors}).
Indeed, the diagonal errors after the first unfolding step for $\lambda_L = 1.5$ are larger than the ones obtained for 
$\lambda_L = 0$, the two being smaller than the ones of the data, due to correlations.

If, after the unfolding, the bin-to-bin correlations enhanced by the transfers of events are too large for the further use
of the spectrum, one could decide to compute this spectrum in less correlated, larger bins (if this would not imply a too 
large lost of information on the shape of the spectrum).
This could be achieved through the use of a different (smaller) number of bins for the $tMC$ distribution or by a rebinning 
operation after the unfolding.

All the functions listed in Eq. (\ref{f:regFunctions}), (\ref{f:1Mexp}) and (\ref{f:1Mfrac}) yield very similar results for 
the unfolding in this example, although the suitable regularization parameters can be different for them.

\section{Conclusions}
In this paper we introduced an unfolding procedure allowing one to deal with a series of problems that one can meet in practice,
and which were not considered by the previous methods.
Using a regularization function, one can consistently compare MC with data spectra, even when the last one contain structures
which were not initially simulated.
It is the same regularization function that allows one to unfold spectra containing important fluctuations from the background
subtraction, without generating biases by the transfer of these events in other regions of the spectrum.
Further more, one can dynamically avoid spurious fluctuations of the method when performing iterations to reconstruct structures
that were not initially simulated.
This method also allows the user to keep under control the bin-to-bin correlations of the final spectrum, while the needed
systematic corrections are performed.

We have described a general unfolding strategy that can be used for rather complicated situations, where all the problems
discussed before are present.
A simplified strategy has been studied and it can be applied for more common problems.
We have described the way one can build reliable toy tests inspired by data, which can provide a good parameterization of the
procedure.
We also propose some tests for the intermediate or final results of the procedure, to be performed directly when unfolding 
the data.

Statistical errors can be directly propagated in this method. 
We have proposed several ways of estimating the systematic effects associated to the method, as well as the ones dues to the 
poor knowledge of transfer matrix.

The C++ source code (relying on ROOT functionality) for the unfolding procedure (and the various described tests) can be obtained
upon request from the author.

Several variations of the unfolding strategy and/or intermediate steps, besides the ones mentioned in the text, are possible.
In the general strategy, one could for example simultaneously compute the estimate of the remaining background fluctuations 
and improve the transfer matrix.
This could be achieved by a straight-forward merging of the formulas \ref{eq:imprbu} and \ref{eq:imprA}, with the use of 
the same $\lambda_S = \lambda_M$ parameter.
One could even place the events from the estimate of the remaining background fluctuations on the diagonal of the transfer
matrix.
In this case, the corresponding events are not moved by the unfolding, just because they are ``predicted'' to be kept in place
by the transfer matrix.
Further on, at the different steps of the general unfolding strategy described in this paper (or of its possible variations),
one could use different functions or parameters, which would yield an even more general method.

\section{Acknowledgments}
The author would like to thank Volker Blobel, Michel Davier, Sebastien Descotes-Genon 
and Andreas H\"ocker for very interesting and useful discussions on these topics, 
and for a careful reading of this manuscript.
This work was supported in part by the EU Contract No.
MRTN-CT-2006-035482, \lq\lq FLAVIAnet''.

%% The Appendices part is started with the command \appendix;
%% appendix sections are then done as normal sections
\appendix

\appendix
\section{Notation}
\label{App:Notation}
As the acceptance correction will be performed at the very end, only events passing 
all the physical cuts will be implicitly considered when computing the quantities in the 
following definitions :
\begin{itemize}
\item
$NB_{d}$ = number of bins in the data and in the reconstructed MC distribution
\item
$NB_{u}$ = number of bins in the unfolding result and in the true MC distribution
\item
$A_{ij}$ = the number of simulated events which were generated in the bin $j$ and 
reconstructed in the bin $i$.
\item
$P_{ij}$ = the probability for an event generated in the bin j to be reconstructed 
in the bin i (folding probability matrix, as estimated from the MC simulation)
\item
$\tilde{P}_{ij}$ = the probability for the ``source'' of an event reconstructed in the bin i to be
situated in the bin j (unfolding probability matrix, as estimated from the MC simulation)
\item
$R$ = rebinning transformation, to pass from data to unfolding bins. 
This transformation can be non-linear.
\item
$R'$ = inverse rebinning transformation, to pass from unfolding to data bins. 
This transformation can be non-linear.
\item
$d_i$ = number of reconstructed data events in the bin i (after background subtraction)
\item
$bd_i$ = number of events in the bin i of the data distribution which are potentially due
to a fluctuation in the subtracted background 
\item
$dk_i$ = number of reconstructed data events in the bin i, which do not correspond to structures 
which are not ``known'' by the MC. 
This number is exactly known in a simulation, but only an estimation of it can be computed
when unfolding real data.
\item
$tdk_j$ = number of true data events, ``known'' by the MC. This distribution can be computed
for toy tests only.
\item
$bias_j$ = events corresponding to structures in the true data, which are not simulated 
in the MC.
\item
$\sigma(d_i)$ = error of $d_i$
\item
$u_j$ = number of unfolded events in the bin j
\item
$bu_j$ = number of events in the bin j of the unfolded distribution which are potentially due
to a fluctuation in the subtracted background
\item
$\sigma(u_j)$ = error of $u_j$
\item
$r_{MCi}$ = number of reconstructed MC events in the bin i
\item
$\sigma(r_{MCi})$ = error of $r_{MCi}$
\item
$t_{MCj}$ = number of true MC events in the bin j
\item
$\sigma(t_{MCj})$ = error of $t_{MCj}$
\item
$NE_d$ = number of events in the data and, by preservation, in the unfolding result
\item
$NE_{mc}$ = number of events in the true and reconstructed MC spectra
\item
$NE_{dSmc}$ = estimate of the number of events in the data, which were simulated in the MC
\item 
$f(\Delta x,\sigma,\lambda)$ = regularization function, 
depending on $\Delta x$, the absolute deviation between two values to be compared, $\sigma$, the 
estimate of the total error to be used for the comparison of the given bin of two spectra,
and $\lambda$, a positive regularization parameter.
\end{itemize}

\end{document}